\documentclass[iop,revtex4]{emulateapj}
\usepackage{lscape}
\usepackage{graphicx}
\usepackage{aas_macros}
\usepackage{amssymb}
\usepackage{calc,float,amsmath,booktabs}
\usepackage{times}
\usepackage{afterpage,listings}
\newcommand       \be        {\begin{equation}}
\newcommand       \ee        {\end{equation}}

\newcommand       \kms		{\,{\rm km \,\, s}^{-1}}

\newcommand       \cm		{\,{\rm cm }}
\newcommand       \msun		{\,{\rm M_\odot}}
\newcommand       \rsun		{\,{\rm R_\odot}}
\newcommand       \lsun		{\,{\rm L_\odot}}
\newcommand       \mspy 		{\,{\rm M_\odot \, yr^{-1}}}
\newcommand       \yr		{\,{\rm yr }}
\newcommand       \s			{\,{\rm s }}
\newcommand       \hr		{\,{\rm hr }}

\newcommand       \K			{\,{\rm K }}
\newcommand       \Hz		{\,{\rm Hz }}

\newcommand       \erg		{\,{\rm erg }}

\newcommand       \AU		{\,{\rm AU}}

\newcommand 		\tg			{t_{\rm g}}

\newcommand 		\theat   	{t_{\rm heat}}
\newcommand 		\teddy   	{t_{\rm eddy}}
\newcommand 		\tcc		   {t_{\rm cc}}
\newcommand       \tl		   {t_{\rm leak}}
\newcommand       \tnu		   {t_{\nu}}
\newcommand 		\tsound   	{t_{\rm sound}}
\newcommand 		\omc			{\omega_{c}}

\newcommand       \omac       {\omega_{ac}}

\newcommand       \M			   {{\mathcal M}_{\rm conv}}
\newcommand 		\vc			{v_{\rm conv}}

\newcommand       \cs		   {c_{\rm s}}

\newcommand       \vgr		   {v_{\rm group, \; r}}
\newcommand 		\vesc			{v_{\rm esc}}

\newcommand 		\Lc				{{\rm L_{conv}}}
\newcommand 		\Lcmax      		{{\rm L_{max, \; conv}}}
\newcommand 		\Lnuc       		{{\rm L_{nuc}}}
\newcommand 		\Lrad       		{{\rm L_{rad}}}
\newcommand 		\Ldamp      		{{\rm L_{damp}}}
\newcommand 		\Ledd       		{{\rm L_{Edd}}}
\newcommand 		\Lnu        		{{\rm L}_{\nu}}
\newcommand 		\Lw				{{\rm L_{wave}}}

\newcommand 		\Ew				{{\rm E_{wave}}}
\newcommand 		\eint			{e_{\rm int}}

\newcommand 		\Mej        		{{\rm M_{ej}}}
\newcommand 		\Rej        		{{\rm R_{ej}}}

\newcommand 		\Mdot       		{\dot{M}}

\newcommand 		\rin        		{r_{\rm in}}
\newcommand 		\rout       		{r_{\rm out}}
\newcommand 		\rprop      		{r_{\rm prop}}

\newcommand 		\rdamp			{r_{\rm damp}}
\newcommand 		\rss     		{r_{\rm ss}}
\newcommand 		\recz       		{r_{\rm env, \; cz}}

\def 	\Lsun		{L$_\odot$}
\def 	\Msun		{M$_\odot$}

\def 	\Zsol		{Z$_\odot$}

\def 	\paperi		{Paper I}

\providecommand		{\adsurl}[1]		{\href{#1}{ADS}}

\newcommand       \brunt			{Brunt-V\"ais\"al\"a}
\newcommand       \MESAstar		{\texttt{MESA star}}

\begin{document}

\shorttitle{Wave-Driven Mass Loss in SN Progenitors} 
\shortauthors{Shiode \& Quataert}

\title{Setting the Stage for Circumstellar Interaction in Core-Collapse Supernovae II:
	Wave-Driven Mass Loss in Supernova Progenitors}

\author{Joshua H. Shiode\altaffilmark{1}}
\affil{Department of Astronomy \& Theoretical Astrophysics Center, University of California, Berkeley, CA 94720-3411, USA}
\altaffiltext{1}{American Astronomical Society, 2000 Florida Ave., NW, Suite 300, Washington, DC 20009-1231, USA}
\email{joshua.shiode@aas.org}

\and

\author{Eliot~Quataert}
\affil{Department of Astronomy \& Theoretical Astrophysics Center, University of California, Berkeley, CA 94720-3411, USA}
\email{eliot@berkeley.edu}





\begin{abstract}

Supernovae (SNe) powered by interaction with circumstellar material provide evidence for intense stellar mass loss during the final years leading up to core collapse. One of the most promising energy sources for powering this mass loss is the prodigious core fusion luminosities during late stages of stellar evolution.   We have argued that during and after core neon burning, internal gravity waves excited by core convection can tap into the core fusion power and transport a super-Eddington energy flux out to the stellar envelope, potentially unbinding up to $\sim \msun$ of material.   In this work, we explore the internal conditions of SN progenitors using the 1-D stellar evolution code \MESAstar, in search of those most susceptible to wave-driven mass loss.   We focus on simple, order of magnitude considerations applicable to a wide range of progenitors.   Wave-driven mass loss during core neon and oxygen fusion happens preferentially in either lower mass ($\lesssim 20 \msun$ ZAMS) stars or massive, sub-solar metallicity stars.  Roughly 20 per cent of the SN progenitors we survey can excite $\sim 10^{46-48} \erg$ of energy in waves that can  potentially drive mass loss within a few months to a decade of core collapse.   This energy can generate a circumstellar environment with $10^{-3} - 1 \msun$ reaching $\sim 100$ AU before explosion.  We predict a correlation between the energy associated with pre-SN mass ejection and the time to core collapse, with the most intense mass loss preferentially happening closer to core collapse.    During silicon burning, a $\lesssim 5$ day long phase for our progenitor models,  wave energy may inflate $\sim 10^{-3} - 1 \msun$ of the stellar envelope to $\sim 10 - 100$s of solar radii. This  suggests that some nominally compact SN progenitors (Type Ibc progenitors) will experience  wave-driven radius inflation during silicon burning and will have a significantly different SN shock breakout signature than traditionally assumed.    We discuss the implications of our results for the core-collapse SN mechanism, Type IIn SNe, Type IIb SNe from extended progenitors (e.g., SNe 1993j and 2011dh), and  observed pre-SN outbursts in SNe 2006jc, 2009ip, and 2010mc.

\end{abstract}



\section{Introduction} \label{sec:intro}

There is a large and growing body of evidence demonstrating that many supernova (SN) progenitors experience episodes of intense mass loss as late as the weeks leading up to core collapse \citep[e.g.,][]{foley2007,ofek2010,ofek2013,mauerhan2013,margutti2013}. These progenitors stand in stark contrast to the canonical picture of massive stars in their final $\sim 1000 \yr$ prior to explosion, that of a ``frozen'' stellar envelope overlying a vigorously burning, neutrino-cooled core \citep[e.g.,][]{kipp,woosley2002}. Direct observations of luminous outbursts that precede supernovae (SNe) like 2006jc \citep{foley2007,pastorello2007}, 2009ip \citep{margutti2013,mauerhan2013} and 2010mc \citep{ofek2013} and the unseen episodes of highly enhanced mass-loss inferred from observations of Type IIn SNe \citep{kiewe2012,gal-yam2012,ginzburg2012} point to dynamic conditions in the envelopes of SN progenitors as they approach explosion. 

The mass-loss rates inferred, assuming SN emission powered by the interaction of the outgoing SN shock and the prior ejected mass, far exceed what is reproducible by line-driven winds, with derived rates of $\sim 10^{-3} - 1 \mspy$ \citep[e.g.,][]{kiewe2012,fox2013,smith2007a}. The total circumstellar masses inferred approach tens of solar masses in the most extreme cases \citep[e.g.,][]{ginzburg2012,moriya2012}. Several candidate mechanisms capable of generating mass loss this prodigious have been proposed, including mass loss driven by waves excited in the convective core \citep[][hereafter \paperi{}]{quataert2012}, hydrodynamic instabilities driven by vigorous convection not well-described by mixing length theory \citep{smith2013}, common envelope interaction with a close companion \citep{chevalier2012,soker2013}, the pulsational pair instability \citep{ppimechanism}, and local radiation-driven instabilities in the stellar envelope \citep{suarez2013}.   The first two of these appear to best explain the large incidence of episodic mass-loss in the final decades leading up to core collapse \citep[e.g.][]{ofek2013}; we focus on the wave-driven mass loss mechanism here.

In this paper, we address two basic questions:  how much energy is available in convectively excited waves during the final evolutionary phases of massive stars? And, in which evolutionary phases can this energy escape the core and generate a pre-SN outburst? To that end, we present an investigation of the interior conditions of a suite of core collapse SN progenitor models, spanning a decade in mass, metallicities from Population III to solar, and initial rotation velocities up to 80 per cent of critical (breakup).   We focus on single star evolution, although we are fully cognizant of the fact that a large fraction of massive stars are in close binaries \citep{sana2012}.

Several aspects of the physics of wave-driven mass loss  are uncertain and limit our ability to make detailed calculations and predictions (see \S \ref{sec:theory} \& \ref{sec:future}).   For this reason, we restrict ourselves to order of magnitude considerations throughout this study.  In particular, we utilize WKB theory in evaluating  wave excitation, damping, and propagation.    Our goal is to identify general trends with stellar progenitor and evolutionary phase that are plausibly robust even given some of the physics uncertainties.

We begin with a summary of the wave-driving mechanism laid out in \paperi{} (\S\ref{sec:theory}) and a description of our grid of 1-D stellar progenitors (\S\ref{sec:mods}). In \S\ref{sec:res}, we present our investigation of SN progenitor interiors and their susceptibility to wave-driven mass loss. We discuss these results and their application to observed systems in \S\ref{sec:disc} and conclude by highlighting directions for future work (\S\ref{sec:future}).


\section{The wave driving mechanism} \label{sec:theory}

The wave-driven mass loss mechanism is discussed in detail in \paperi{}. In the following, we provide a brief summary of the key concepts, and update some of the relevant timescale considerations.
\subsection{Convective wave excitation} \label{ssec:excitation}

In the final stages of massive stellar evolution, i.e., carbon fusion and beyond, the core of the star is cooled predominantly by thermal neutrinos \citep[e.g.,][]{clayton1984,woosley2002}. The nuclear luminosity is thus in equilibrium with the neutrino cooling, which may exceed the emergent stellar luminosity by many orders of magnitude. If each core burning phase releases $\sim 10^{51} \erg$ by fusing $\sim \msun$ of material, we expect characteristic fusion and neutrino luminosities of $10^{7}$ \Lsun{} for C-burning, $10^{10}$ \Lsun{} for Ne and O-burning, and $10^{12}$ \Lsun{} for Si-burning \citep[based on burning timescales given in Table~\ref{tab:mods}, and e.g.,][]{woosley2002}. In most models, the luminosities during C and Ne-fusion are in fact smaller by a factor of $\sim 10$ due to the small abundances of C and Ne left behind by the prior burning phases. The emergent radiative luminosity during these phases is roughly equal to the Eddington luminosity for electron scattering, $\sim 10^{5} - 10^{6}$ \Lsun{} for core collapse SN progenitors. 

Due to the different temperature dependences of the nuclear burning and neutrino emission rates, local regions within the core are convectively unstable, with convection carrying a significant fraction of the nuclear luminosity, $\lesssim 10$ per cent. Thus, convection may carry a luminosity that exceeds the envelope Eddington value during C-burning, and does so by many orders of magnitude during Ne burning and beyond. This convection has characteristic mach numbers of $\sim 10^{-3} - 0.03$.\footnote{The stellar evolutionary calculations in this paper generally predict Mach numbers a factor of few smaller than the rough analytic estimates given in Table 1 of \paperi{}.}

\begin{figure*}
   \centering
   \includegraphics[width=0.9\textwidth]{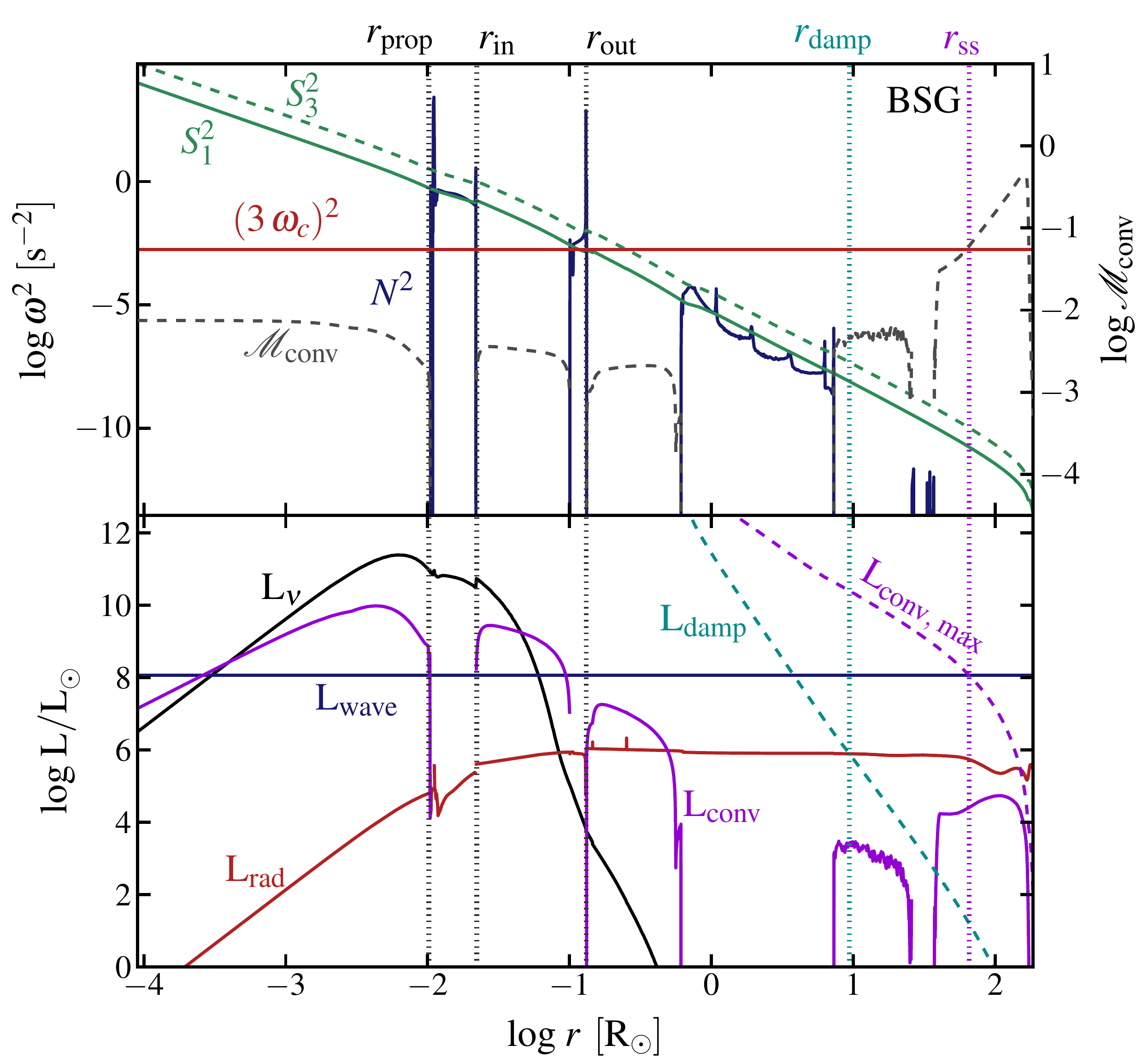}
   \caption{{\em Top:} Propagation diagram for a 40 \Msun{} $10^{-1}$ \Zsol{} blue supergiant progenitor at core oxygen burning. \brunt{} frequency (solid blue line), Lamb frequency for $\ell = 1, 3$ (solid and dashed green lines, respectively) and the propagating g-mode frequency excited by  core convection  (red solid line) are shown against the left axis. The convective mach number is plotted as the dashed grey line against the right axis.  {\em Bottom:}  Luminosities relevant for wave excitation and damping are shown in each shell of the model: the neutrino luminosity ($\Lnu$) in solid black, radiative ($\Lrad$) in solid red, convective ($\Lc$) in solid magenta, the radiative damping luminosity ($\Ldamp$) in dashed teal (see eqn.~\ref{eqn:ldamp}), maximum convective luminosity ($\Lcmax$) in dashed magenta (see eqn.~\ref{eqn:lbreak}), and the wave luminosity ($\Lw$) excited during this phase as the blue solid line. The radii relevant for wave-driven mass loss are marked by vertical dotted lines, and labeled at the top. The propagation radius, $\rprop$, where the internal gravity waves excited by core convection start to propagate is shown in grey, as are the inner and outer radii of the tunneling region, $\rin$ and $\rout$, respectively. The damping radius, $\rdamp$, is shown in teal, where $\Lrad = \Ldamp$; $\rdamp$ is where the acoustic waves in the envelope damp by radiative diffusion.  $\rss$ is shown in magenta, where $\Lw = \Lcmax$; $\rss$ is where the convection driven by wave energy deposition becomes supersonic and likely initiates an outflow. }
   \label{fig:radii}
\end{figure*}

At the interfaces between these convection zones and neighboring stable regions, a radius we call $\rprop$, convection transfers a fraction of its luminosity to linear, propagating gravity modes (g-modes) in that layer \citep[e.g.,][]{press1981,goldreich1990,lecoanet2013}.   This has been seen explicitly in  \citet{meakin2006}'s hydrodynamic simulations of the late stages of massive stellar evolution.      The top panel of Fig.~\ref{fig:radii} shows an example propagation diagram  for a 40 \Msun{} $10^{-1}$ \Zsol{} progenitor during core oxygen burning; here $\rprop$ is the innermost labeled radius, marking the boundary between the oxygen burning core convection zone and the neighboring stably stratified layer where g-modes propagate (in which the square of the \brunt{} frequency, $N^2$, is positive).

In \paperi{}, we used the estimate from the earlier literature that the energy flux in internal gravity waves is $\Lw \sim \M \, \Lc$, where $\Lc$ is the convective luminosity and $\M$ is the convective mach number of the energy-bearing eddies (eqn.~1 from \paperi{}). This is, however, only appropriate for a model in which the transition between the convective and radiative zones is discontinuous \citep{lecoanet2013}. High frequency internal gravity waves have longer wavelengths in the radiative zone and thus indeed see the transition as approximately discontinuous. This is not, however, the case for the gravity modes that carry most of the wave energy flux, those with frequencies comparable to the convective turnover frequency (which have shorter wavelengths in the radiative zone). The excitation of these energy-bearing waves depends on the structure of the radiative-convective transition and thus depends on details that are not well-modeled in 1-D stellar evolution calculations. \citet{lecoanet2013} argued that a smooth transition is more physical and leads to a larger energy flux than we assumed in \paperi{}. For a particular analytically tractable smooth model of the radiative-convective transition, they found that the wave luminosity excited by convection is

\be \label{eqn:lwave}
   	\Lw \sim \M^{5/8} \, \Lc \sim 10^{8} \left(\frac{\Lc}{10^{10} L_\odot}\right) 
		\left(\frac{\M}{10^{-3}} \right)^{5/8} \, {\rm L}_\odot,
\ee
where there is a dimensionless pre-factor in front of eqn.~\ref{eqn:lwave} that depends on the properties of the radiative-convective transition. It is not possible to calculate this pre-factor with existing 1-D stellar evolution calculations. Thus we set this to 1 while noting that it could be smaller if the radiative-convective transition is very thin (i.e., closer to discontinuous).   


\begin{figure}
   \centering
   \includegraphics[width=\columnwidth]{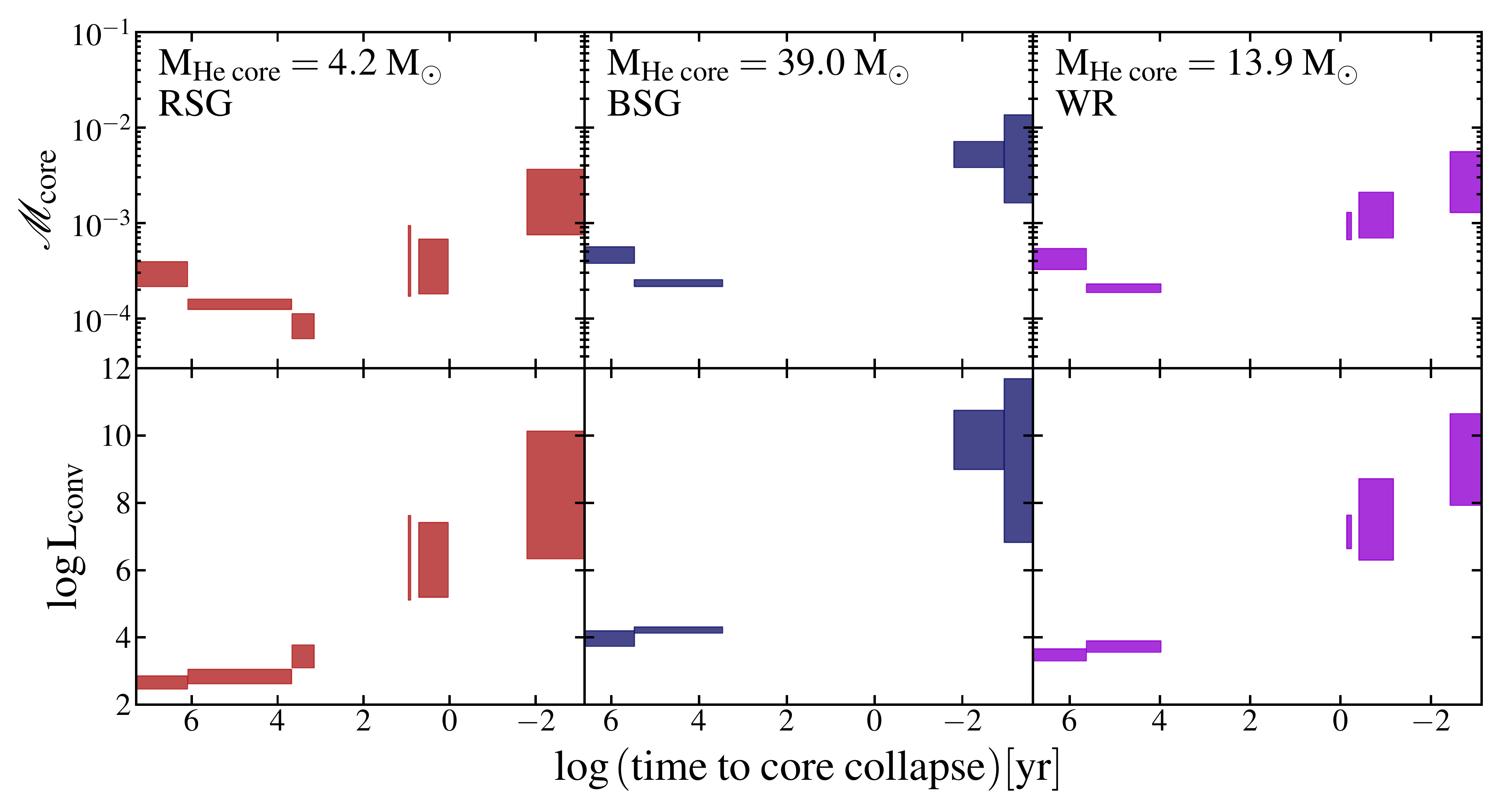}
   \caption{Volume-averaged core convective mach number ({\em top panels}) and convective luminosity ({\em bottom panels}) as a function of time to core-collapse for three example stellar models (one of each of the three progenitor classes defined in \S \ref{sec:mods}; each is labeled by He core mass at core-collapse).   The range of values on the x axis corresponds to the duration of the burning phase while the range of values on the y axis corresponds to the 95th percentile during the burning phase.   Results are shown from core hydrogen to core silicon burning, for each phase in which core fusion proceeds convectively (core carbon and neon burning are radiative in some models).   The convective Mach number and luminosity peak during late stages of stellar evolution, significantly increasing the energy flux in  gravity waves (eq. \ref{eqn:lwave}). }
   \label{fig:mach}
\end{figure}


The spectrum of waves excited by convection is also uncertain (though constraints may be forthcoming from precision observations of massive main sequence stars; see \citealt{shiode2013}). In broad terms, the convection most efficiently excites waves with characteristics similar to the energy-bearing eddies \citep[e.g.][]{press1981}. Thus, the wave energy is likely concentrated in waves with characteristics similar to the energy-bearing eddies in the convection zone. These have frequencies near $\omc$, where $\omc \approx \vc / \min (r, H)$ is the \brunt{} frequency in the convection zone, $r$ is the radius coordinate, $\vc$ is the convective velocity and $H$ is the pressure scale height; the horizontal length-scales correspond to spherical harmonic degrees $\ell \sim \rprop / H(\rprop)$. 

The lowest frequency waves, those with $\omega \approx \omc$, have non-linear amplitudes and break as soon as they reach the radiative zone (eqn.~\ref{eqn:lwave} gives the luminosity in linear g-modes and does not include the energy associated with these immediately non-linear waves).\footnote{Note that equation 6 in \paperi{} for the nonlinearity of g-modes has a typo.  The factor $(N/\omega)^{3/2}$ should be $(N/\omega)^{1/2}$.  The correct expression was used for the estimates given in \paperi{}.} Thus, in the following, we focus on the linear, propagating waves with frequencies $\omega \gtrsim 3\, \omc$.   The g-modes are most prone to breaking near the (first) radiative-convective boundary; nonlinear damping  is thus not likely to be a significant damping mechanism at larger radii, at least until the g-modes convert into acoustic waves in the stellar envelope.

Figure \ref{fig:mach} shows the core convective Mach number and convective luminosity as a function of time until core-collapse for three different example stellar models described in more detail in \S \ref{sec:mods} and Table \ref{tab:mods}.   The plot covers all of stellar evolution, from core hydrogen to core silicon burning.   The convective Mach number and convective luminosities peak during late stages of stellar evolution, significantly increasing the energy flux in internal gravity waves predicted by equation \ref{eqn:lwave}.

\vspace{1cm}
\subsection{The fate of gravity waves: tunneling vs. neutrino losses} \label{ssec:gmodefate}

Convectively excited g-modes in the cores of evolved massive stars damp through one of two main channels: locally via neutrino losses or by tunneling out of the g-mode propagation cavity (we will also refer to the latter as ``leakage''). Due to the highly temperature sensitive neutrino emission rates, positive (negative) temperature perturbations associated with waves lose more (less) energy via neutrino emission than the background, damping the perturbations. Nuclear fusion in the core may provide driving via an analogous, but opposite in sign, process known as the $\epsilon$-mechanism \citep{murphy2004}, but we ignore this  contribution for simplicity.  This is conservative in that we likely overestimate the wave damping rate.   Leakage arises because the envelope of the star can often host acoustic waves of the same frequencies as the g-modes propagating in the core. Convectively excited g-modes can couple to the acoustic propagation cavity if they can tunnel through the intervening forbidden region (barrier) and are above the envelope's acoustic cutoff frequency, $\omac \approx \cs / (2\, H)$, where $\cs$ is the sound speed in the stellar envelope \citep[see e.g.,][]{unno}. 

Larger length-scale (low $\ell$) waves decay slower through the barrier and higher frequency waves see smaller barriers (see \paperi{}). Thus, g-modes with these characteristics have the highest probability of tunneling, rather than damping to neutrinos. Whether a progenitor can drive mass loss by convectively excited waves depends first on the capacity for advanced convective burning to excite sufficiently high frequency and low $\ell$ waves to high energy.

G-modes damp primarily via tunneling if they leak faster than they damp to neutrinos:
\be \label{eqn:cond1}
   \tl < \tnu, 
\ee
where the leakage time, $\tl$, is given by the group travel time across the g-mode cavity, divided by the tunneling probability\footnote{In some cases, particularly in compact progenitors, the most energetic, propagating g-modes have frequencies on the same order as the acoustic cutoff frequency in the envelope; there is likely an additional inhibiting contribution to the tunneling probability in these cases because the outer turning point $\rout$ is part of the thin stellar atmosphere. This correction is not accounted for here.}:
\be \label{eqn:tleak}
	\tl \approx \tg \left(\frac{\rout}{\rin}\right)^{2 \, \Lambda},
\ee
and the group travel time is given by
\be \label{eqn:tgroup}
	\tg = \int_{\rprop}^{\rin} \frac{dr}{\vgr},
\ee
where $\Lambda^{2} \equiv \ell (\ell + 1)$, $\vgr$ is the g-mode group velocity in the radial direction, $\rin$ is the inner radius of the tunneling region (outer edge of the g-mode propagation cavity) and $\rout$ is the outer radius of the tunneling region (inner edge of the p-mode propagation cavity; see Fig.~\ref{fig:radii} for the locations of these radii for an example mode). The neutrino damping time is set by the characteristic time for neutrino losses in the g-mode propagation cavity, enhanced by the temperature sensitivity of the neutrino losses:
\be \label{eqn:tnu}
	\tnu \approx \frac{\int_{\rprop}^{\rin} \eint \, dm}
		{\left(\frac{d\ln \epsilon_{\nu}}{d\ln T}\right)_{\rho} \int_{\rprop}^{\rin} \epsilon_{\nu}(r) \,dm},
\ee
where $\eint$ is the internal energy per unit mass for the stellar material, $\epsilon_{\nu}(r)$ is the neutrino energy loss rate per unit mass, $dm \equiv 4 \pi r^{2} \rho \, dr$, and $({d\ln \epsilon_{\nu}} / {d\ln T})_{\rho} \sim 9$ for neutrino losses due to pair-annihilation. 

The size of the g-mode propagation cavity, $\rprop$ to $\rin$, is roughly independent of frequency for the frequencies of convectively excited g-modes (see Fig.~\ref{fig:radii}), fixing the approximate neutrino damping timescale given in eqn.~\ref{eqn:tnu}. The leakage timescale, on the other hand, depends strongly on the width of the tunneling region, $\rout / \rin$ and the rate at which waves decay through the barrier. Both decline with decreasing $\ell$, while the former also declines with increasing wave frequency.

Equation~\ref{eqn:tleak} was derived under the WKB assumption, wherein the waves are assumed to vary on lengthscales much shorter than the background. For many of the compact progenitor models we survey here (i.e. WR stars), this assumption is invalid in the envelope (acoustic propagation cavity) for the frequencies of convectively excited g-modes. Equivalently, these modes have frequencies below the acoustic cutoff for the stellar envelope and are likely reflected before reaching that cavity. In the following, we consider waves with $3\, \omc \lesssim \omac$ as precluded from tunneling. 

There may be an additional suppression of the tunneling probability due to  composition barriers present at the interfaces of convective and radiative zones, which can introduce variations on lengthscales shorter than the wavelength. The width and magnitude of these barriers depends sensitively on the treatment of mixing and convective boundaries. We do not attempt to account for these here, but acknowledge that it is an uncertainty in our calculation.  

\subsection{Acoustic waves and mass loss} \label{ssec:acoustic}

Waves that satisfy eqn.~\ref{eqn:cond1} and have frequencies above the acoustic cutoff frequency in the envelope  tunnel out to the stellar envelope, where they may be carrying a significantly super-Eddington luminosity as acoustic waves. In the envelope, these waves damp either when the radiative damping timescale becomes comparable to the wave travel time or when the wave reaches non-linear amplitudes: $\xi_{r}\, \omega \sim \cs$ (see \paperi{}). The former condition can be represented as
\be \label{eqn:ldamp}
   \Lrad \gtrsim \Ldamp \equiv \frac{4 \pi r^2 \rho \, \cs^3}{(k\, H)^2},
\ee
where $\rho$ is the mass density and $k$ is the wavenumber of the propagating acoustic wave. By contrast, sound waves reach non-linear amplitudes when
\be \label{eqn:lbreak}
   \Lw \gtrsim \Lcmax \equiv 4 \pi r^2 \rho \,\cs^3.
\ee
In the following, we call $\rdamp$ the location where waves satisfy eqn.~\ref{eqn:ldamp} and $\rss$ where they satisfy eqn.~\ref{eqn:lbreak} (where the ``ss'' is short for ``supersonic''). The bottom panel of Fig.~\ref{fig:radii} demonstrates the locations of these radii, based on eqns.~\ref{eqn:ldamp} and \ref{eqn:lbreak}, for an example mode.

Except in some compact progenitors, waves reach eqn.~\ref{eqn:ldamp} first, deeper in the star, after traveling on the sound crossing time, $\tsound$, from $\rout$ to $\rdamp$. The sound crossing time is generally short enough to be unimportant. At $\rdamp$, the wave energy damps by radiative diffusion and creates a local region with a significantly super-Eddington flux. In order to drive an outflow, the deposited wave energy must at a minimum reach $\rss$, where even $\M \gtrsim 1$ cannot carry the total luminosity. If $\rdamp$ is in an envelope convection zone, energy moves quickly outward on a timescale 
\be \label{eqn:teddy}
   \teddy \approx \left.\frac{H}{\vc}\right|_{\rdamp} \approx H(\rdamp) \left(\frac{\Lw}{4 \pi \rdamp^2 \, \rho(\rdamp)}\right)^{-1/3}.
\ee 
If, instead, $\rdamp$ is in a stably stratified radiative zone, the deposited energy must heat the local material enough to drive convection (see, e.g., \citealt{piro2008} for a detailed discussion of the physics of time-dependent convection zones in the context of Type Ia SN progenitors). This occurs on a (generally much longer) timescale 
\be \label{eqn:theat}
	\theat(r) \approx \frac{\int_{\rdamp}^{r} \eint dm}{\Lw},
\ee
where the terminal radius, $r = \min(\rss, \recz)$, and $\recz$ is the base of a pre-existing envelope convection zone (if one exists). If there is a pre-existing convection zone, the wave energy need only heat enough material to extend the convection zone down to $\rdamp$. If there is not one, or $\rss < \recz$, the wave energy must heat all the material between $\rdamp$ and $\rss$ to potentially drive an outflow. This of course depends sensitively on the details of how the convection zone grows; it might in fact grow more quickly by entrainment \citep[as described in e.g.,][]{arnett2009}. We use eqn.~\ref{eqn:theat} in the following as it represents a conservative estimate of the timescale for energy to reach $\rss$ after damping at $\rdamp$. 

\begin{figure}
   \centering
   \includegraphics[width=\columnwidth]{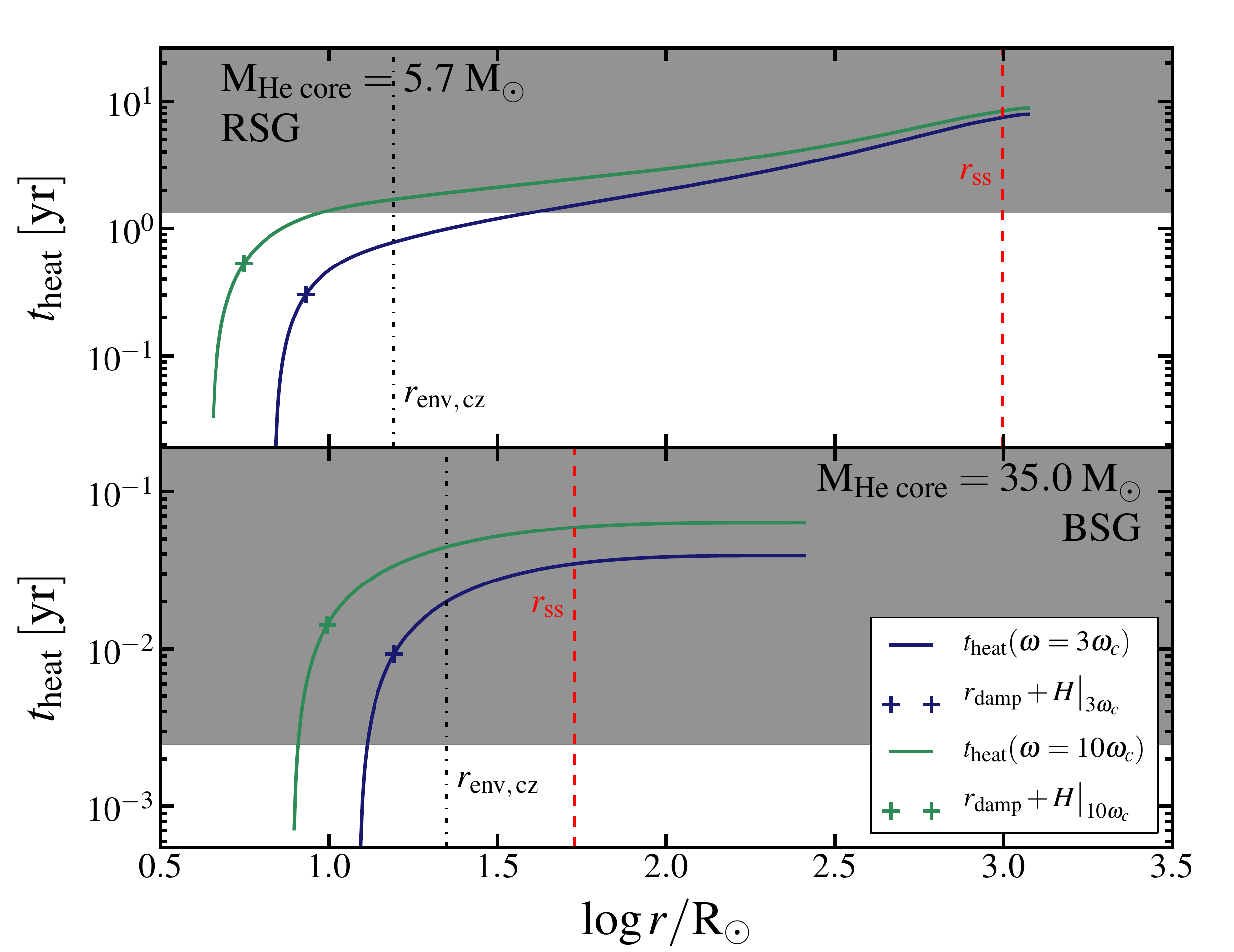}
   \caption{Timescale for wave damping to heat the stellar envelope by a factor of order unity ($\theat$; eq. \ref{eqn:theat}) out to a given radius $r$, for  waves with frequencies of 3 and 10 $\omc$ (blue and green lines, respectively).  Crosses mark one scale height above the radius where the waves initially damp ($\rdamp$). The dash-dot black vertical lines show the radius of the base of the pre-existing envelope convection zone and the dashed red line shows the sonic point, $\rss$, for a potential outflow  (see eqn. \ref{eqn:mdot}).  Once waves can heat the stellar envelope out to $\sim \recz$, convection rapidly carries the energy out on a timescale $\ll \theat$.   The grey span shows the region in which the heating time is longer than the time to core collapse and there is thus not sufficient time for the wave energy to significantly modify the stellar structure.   {\em Top panel:}  waves are able to heat the envelope out to $\recz$ prior to core collapse, thus likely powering an outflow. $\Lw = 5 \times 10^5 \lsun$ at core oxygen fusion for this low He core mass model.   {\em Bottom panel:}  $\Lw = 3 \times 10^9 \lsun$ at core oxygen fusion for this high He core mass model. Despite the much higher wave luminosity, the heating timescale is nonetheless longer than the time to core collapse for this more massive and compact progenitor and an outflow is unlikely. \label{fig:theat}} 
\end{figure}

To illustrate the importance of the heating timescale in determining the effect of the dissipated wave energy, Figure \ref{fig:theat} shows $\theat$ as a function of radius for two models during core oxygen fusion.   In the upper panel, the timescale to heat the stellar envelope out to the pre-existing envelope convection zone ($\recz$; vertical dash-dot line) is less than the time to core collapse (shown by the shaded grey region) and thus the dissipated wave energy can be efficiently carried to the stellar surface where it can plausibly power an outflow.  In the lower panel, however, the timescale to heat the envelope out to the pre-existing envelope convection zone is significantly longer than the time to core-collapse, likely precluding the wave energy from having a significant effect on the stellar structure prior to core collapse.

Convection necessarily fails to carry the energy at $\rss$, where the total luminosity exceeds $\Lcmax$. At this point, even $\M \gtrsim 1$ convection cannot carry the total luminosity. This can result in a strong pre-SN outflow, so long as $\tsound$, $\theat$ and $\teddy$ are all shorter than the time to core collapse. In compact progenitors where convectively excited waves have frequencies above the envelope acoustic cutoff, waves sometimes reach eqn.~\ref{eqn:lbreak} first.  In this case, the heating timescale is invariably quite short and the wave energy potentially drives a strong outflow on roughly the sound travel time to the stellar surface.


For progenitors in which wave energy can reach $\rss$ prior to core collapse, we assume that this leads to an outflow and estimate the mass loss rates and total potential ejecta mass for the given burning phase as follows. At $\rss$, $\M \gtrsim 1$ convection is required to carry the outgoing flux; thus $\rss$ represents a sonic point in a potential outflow. The mass loss rate for this outflow will be 
\be \label{eqn:mdot}
   \Mdot \approx 4 \pi \rss^2 \, \rho \, \cs(\rss).
\ee
Since $\Lw = \Lcmax$ at $\rss$, the kinetic energy of the proposed outflow, $1/2 \,\Mdot \vesc^2 (\rss)$, will exceed the wave luminosity so long as $\cs < \vesc$, which it must be for bound stellar material.   Thus, the outgoing wave luminosity cannot energetically sustain an outflow with the full $\Mdot$ implied by the sonic point arguments; the star is in a regime analogous to the ``photon-tired'' wind discussed by \citet{owocki2004}. This could lead to inflation of the envelope to larger radii (as argued by \citealt{soker2013}). However, we view this as unlikely because the total wave energy deposited at $\rss$ generally exceeds the binding energy of the star at $r > \rss$. Thus, we suspect that an outflow with a smaller $\Mdot$ than given in equation \ref{eqn:mdot} is the most likely outcome. \citet{owocki2004} also argue for the latter, positing the development of a porous atmosphere in the presence of super-Eddington luminosities. This provides low density channels out of which radiation can flow and drive a lower density wind. The true outcome of this super-Eddington energy deposition clearly depends on multi-dimensional effects, which will need to be examined using hydrodynamical models. These are, however, beyond the scope of this work.

In this ``photon-tired" regime, we very roughly estimate the ejecta mass by assuming that the wind taps into the full wave luminosity via a lower $\Mdot$ outflow:
\be \label{eqn:Mej}
   \Mej \lesssim \frac{2 \, \Ew}{\vesc^2 (\rss)}.
\ee
The corresponding mass loss rate is
\be \label{eqn:mdotlim}
   \Mdot \approx \frac{\Mej}{t_{\rm fusion}},
\ee
where $t_{\rm fusion}$ is the timescale for the relevant burning phase that generates $\Ew$ used in eqn.~\ref{eqn:Mej}.    We note that our analysis more robustly predicts $\Ew$ than either $\Mej$ or $\Mdot$ since the latter depend on both how the wave energy powers an outflow and the associated speed of the outflow.


\section{Stellar Models} \label{sec:mods}

We use the \MESAstar{}\footnote{http://mesa.sourceforge.net/} stellar evolution code \citep[version {\bf 4789};][]{mesa2011} to construct evolutionary sequences from the zero-age main sequence (ZAMS) to core collapse for SN progenitors ranging in initial mass from 12 to 100 \Msun{}, metallicity from 0 (i.e., Population III) to solar, and initial angular velocities from 0 to 0.8 critical (see Appendix A for more details on the \MESAstar{} parameters we employ). \citet{mesaII} provides updated, detailed descriptions of the stellar evolution physics and numerical scheme employed by \MESAstar{}. Thus, we provide only a brief summary of the key aspects of our specific calculations, and refer the reader to that comprehensive work for details.

All models employ the \citet[]{gs98} chemical mixture, OPAL opacities \citep{opal}, and updated nuclear reaction parameters for the $^{12}{\rm C(\alpha, \gamma)^{16}O}$ \citep{kunz2002}, $^{14}{\rm N(p, \gamma)^{15}O}$ \citep{imbriani2005}, and triple-alpha \citep{fynbo2005} reactions. 

We determine convective boundaries using the Ledoux criterion, with a mixing length parameter $\alpha = 1.5$, semiconvection with a dimensionless efficiency parameter, $\alpha_{sc} = 0.1$, and thermohaline mixing with efficiency, $\alpha_{th} = 2.0$. On the main sequence, we use 33.5 per cent of a pressure scale height of overshoot above the convective core\footnote{We use a step function overshoot prescription, in which the convection zone is extended a distance of 33.5 per cent of a pressure scale height above the Ledoux boundary, with a constant diffusion coefficient.}, following the results of \citet{brott2011}. Beyond the main sequence, we ignore overshoot for simplicity.

For the majority of our calculations, we assume the theoretical mass loss rates of \citet*{vink2001} when $T_{\rm eff} > 10^{4} \K$ and \citet{de-jager1988} when $T_{\rm eff} < 10^{4} \K$, each scaled down by a dimensionless efficiency factor of $0.8$ for non-rotating models (``0.8~(v+dj)'') and $0.6$ for rotating models (``0.6~(v+dj)''). For rotating models, the mass-loss rate is allowed to increase above this prescription by up to a factor of ten, in order to expel any surface layers whose rotation would be super-critical (see \citealt{mesaII} for a more thorough description). In order to test the effect of the mass loss prescription on the results, we also include calculations for high mass, solar metallicity progenitors, where we use the \citet{nieu1990} rates for $T_{\rm eff} < 10^{4} \K$ and a dimensionless efficiency factor of $1.0$ (``v+n''); this more closely matches the prescription described in \citet{woosley2002}. 

For rotating progenitors, we employ compositional mixing and angular momentum transport via all magnetohydrodynamic (MHD) instabilities available in version 4789 of \MESAstar{}, except that we follow \citet{mesaII} and ignore the Solberg-Hoiland contribution to the convective instability criterion. The magnetic field is assumed to arise via the Spruit-Tayler dynamo \citep{spruit2002}. In detail, we do not believe that the angular momentum transport processes in stellar interiors are well understood. We include the rotating progenitor models to give an order-of-magnitude indication of how different single star evolution outcomes can be when stellar rotation is included.


\begin{figure}
   \centering
   \includegraphics[width=\columnwidth]{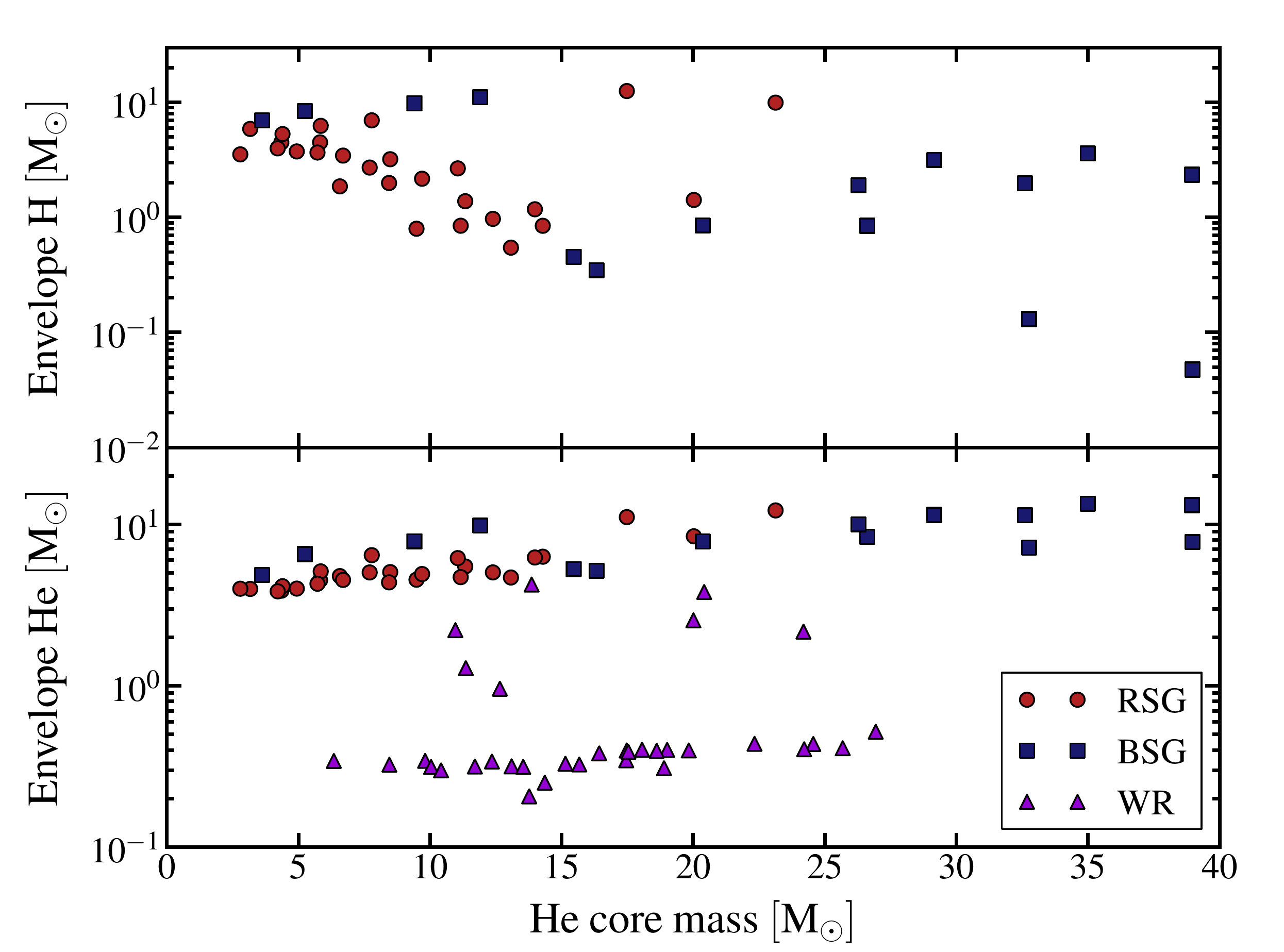}
   \caption{Total masses of hydrogen ({\em top}) and helium ({\em bottom}) in the envelopes of our SN progenitor models, as a function of He core mass.   Envelope masses are based on the final stellar model near the end of silicon fusion.  Classifications into blue and red supergiants and Wolf-Rayet stars are as described in \S \ref{sec:mods}.  Note that all WR models have lost their hydrogen envelope and would thus produce Type Ibc SNe.} 
   \label{fig:envmass}
\end{figure}



\begin{figure*}
      \centering
      \includegraphics[width=0.9\textwidth]{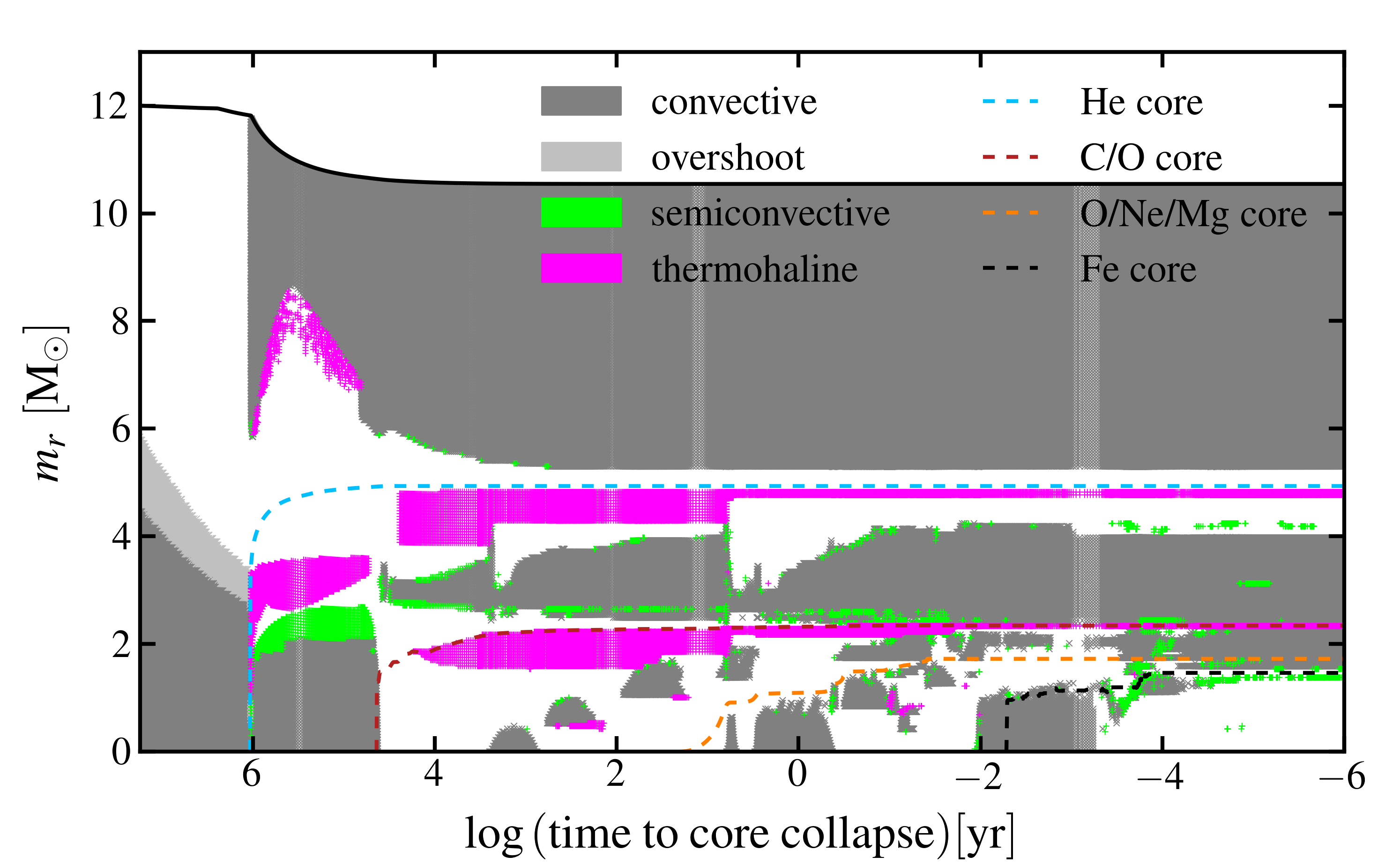}
      \caption{The convective history of a 12 \Msun{}, solar metallicity star initially rotating at 20 per cent of the critical rotation velocity, plotted against log time to core collapse ($\tcc$), in years. This star becomes a red supergiant SN progenitor with a 4.94 \Msun{} helium core and a 5.61 \Msun{} hydrogen envelope. For each timestep, the dark grey shaded region represents mass coordinates that are convective, light grey those that have overshoot, green those with semiconvection, and magenta those with thermohaline mixing. For example, for $\log \tcc \gtrsim 6$ (far left), the star is on the main sequence and has a convective core burning hydrogen that recedes in mass with time from $m_r \sim 4.5 - 3 \msun$, with a $\sim 1.5 \msun$ overshoot region above it. The dashed lines show the mass boundaries of the various cores, light blue shows the boundary of the helium (i.e., hydrogen-depleted) core, red the C/O (i.e., H and He-depleted) core, orange the O/Ne/Mg core, and black the iron core. The O/Ne/Mg core, for example, moves out in mass with time as carbon is depleted by burning in the core and shells.}
      \label{fig:kipp12}
\end{figure*}
   


\begin{figure*}
      \centering
      \includegraphics[width=0.9\textwidth]{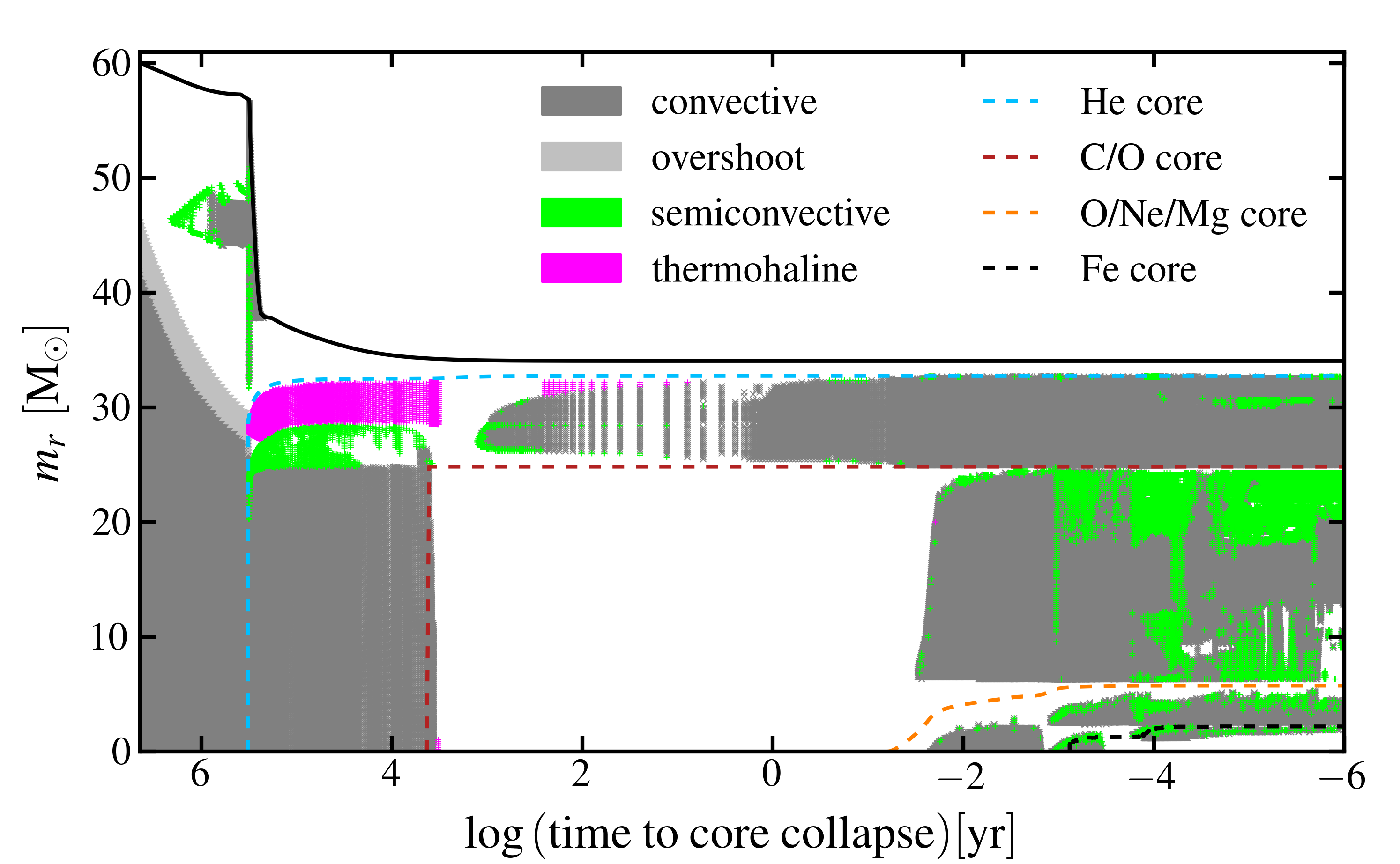}
      \caption{The convective history of a 60 \Msun{}, one-tenth solar metallicity non-rotating star, plotted against log time to core collapse, in years. This star becomes a blue supergiant SN progenitor with a 32.75 \Msun{} helium core and a 1.3 \Msun{} radiative hydrogen envelope. Colors and lines as in Fig.~\ref{fig:kipp12}.}
      \label{fig:kipp60}
\end{figure*}



\begin{figure*}
      \centering
      \includegraphics[width=0.9\textwidth]{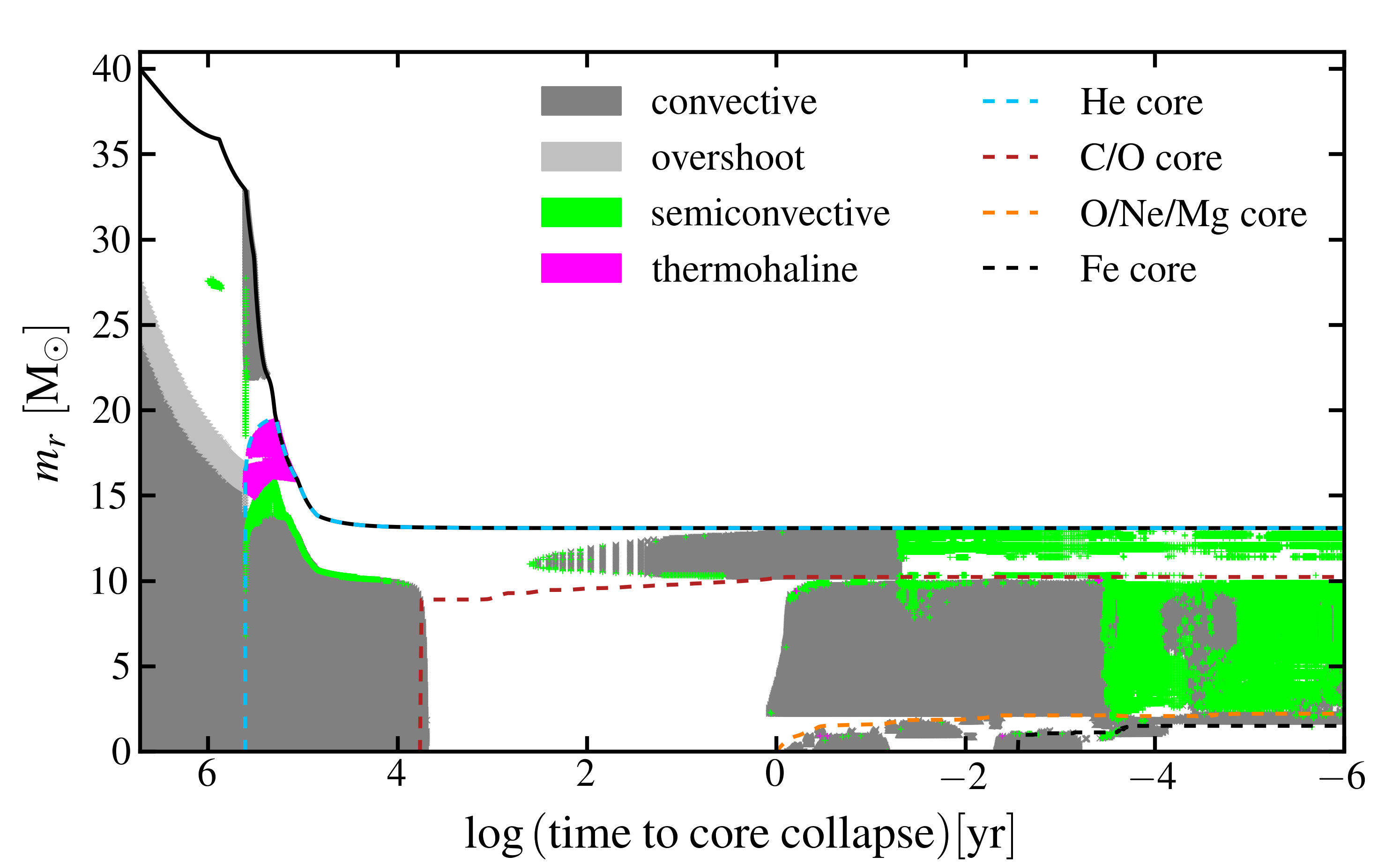}
      \caption{The convective history of a 40 \Msun{}, solar metallicity, non-rotating star, plotted against log time to core collapse, in years. This star becomes a Wolf-Rayet SN progenitor with a 13.1 \Msun{} helium core. Colors and lines as in Fig.~\ref{fig:kipp12}.}
      \label{fig:kipp40}
\end{figure*}


Table \ref{tab:mods} in Appendix B summarizes the properties of our grid of evolutionary calculations for a total of 76 progenitor models. The table gives the initial values for mass, metallicity, and rotation, as well as the mass loss scheme, and shows how these map to final core masses, burning timescales, and our progenitor classifications. Progenitor classifications are based on the structure of the final model prior to core collapse:
\begin{itemize}
	\item{Red supergiant (RSG):  hydrogen envelope and $T_{\rm eff} < 10^{4} \K$.}
	\item{Blue supergiant (BSG):  hydrogen envelope and $T_{\rm eff} \geq 10^{4} \K$.} 
	\item{Wolf-Rayet (WR):  a star that has lost its hydrogen envelope.}
\end{itemize}
These classifications correspond to stars with final radii $300 - 10^3 \rsun$ for RSGs, $3 - 300 \rsun$ for BSGs and $0.4 - 1.6 \rsun$ for WRs.    The hydrogen and helium envelope masses for all of our progenitors are shown in Figure \ref{fig:envmass} as a function of helium core mass just prior to core collapse (we define the He core mass to be the total stellar mass within the mass coordinate where the hydrogen mass fraction first drops below $10^{-4}$).    Although none of the progenitors lose their helium envelopes prior to core collapse, several of our WR progenitors have sufficiently low helium envelope masses that they might produce Type Ic SNe upon explosion (based on \citealt{hachinger2012}).   

Broadly, we find that our solar metallicity progenitors with ${\rm M}_{\rm ZAMS} \gtrsim 30 \msun$ become WR stars, lower ZAMS mass solar metallicity stars generally produce RSGs, and low-metallicity stars produce BSGs. At fixed ${\rm M}_{\rm ZAMS}$, moderate rotation produces bigger helium core masses through mixing, while the largest initial rotation values lead to smaller cores due to the rotational enhancement of the stellar wind.

Figures~\ref{fig:kipp12}, \ref{fig:kipp60} and \ref{fig:kipp40} show examples of the convective history, plotted against $\log \tcc$, where $\tcc$ is time to core collapse, for each of our progenitor categories. Fig.~\ref{fig:kipp12} shows the history of a 12 \Msun{}, solar metallicity, slowly rotating progenitor that becomes a RSG; Fig.~\ref{fig:kipp60} shows a 60 \Msun{}, 0.1 \Zsol{}, non-rotating progenitor that becomes a BSG; Fig.~\ref{fig:kipp40} shows a 40 \Msun{}, solar metallicity, non-rotating progenitor that makes a WR star. In these diagrams, stellar mass is plotted against the time to core collapse, in years, so that each plot shows the convective history from the ZAMS to $\sim 30$ seconds from core collapse ($\log \tcc = -6$). The mass extent of each convection zone in the star is marked by the darker gray shaded regions, overshoot mixing (on the main sequence only) is shown in light gray, semiconvection in green, and thermohaline mixing in magenta. Also plotted are the mass boundaries of the He, C/O, O/Ne/Mg, and iron cores as a function of time to core collapse. For the He, C/O and O/Ne/Mg cores, these boundaries are determined by the first mass coordinate interior to the surface (or the previous mass boundary) where hydrogen, helium or carbon drops below a mass fraction of $10^{-4}$. For the iron core, the boundary is determined by the first mass coordinate inward from the boundary of the O/Ne/Mg core where the total mass fraction of iron-group elements rises above 0.5.

In massive stellar envelopes, radiation pressure provides a significant fraction of the pressure support even when the opacity is set by electron-scattering \citep{kipp}. Significant opacity enhancements due to transitions of iron at $\log T \sim 5.2$ and the recombination of HeII ($\log T \sim 4.7$) and H ($\log T \sim 4$) lead to regions of the stellar envelope where the Eddington luminosity is locally exceeded and gas pressure and density inversions can result \citep{mesaII}. In order to evolve through these computationally difficult stages of evolution, we employ the ``enhanced mixing length theory'' described in \citet{mesaII}, which may crudely account for an energy transport mechanism not present in our 1-D models, such as the development of porosity \citep{shaviv2001,owocki2004}. We thus sacrifice accuracy in the surface evolution in favor of completing the core evolution all the way to core collapse. This implies, however, that our progenitor classifications (RSG, BSG, and WR) have significant uncertainties (as do, however, all analogous results in the literature!).


\section{Results} \label{sec:res}

We present the results of our calculations by core burning phase: carbon, neon, oxygen, and silicon burning. Since burning timescales can vary by more than an order of magnitude, decreasing for increasing core mass, this does not uniformly map onto time to core collapse for all progenitors (see \S\ref{sec:disc}).

In order for convectively excited waves to drive mass-loss in a SN progenitor at any given time, they must satisfy the following conditions:
\begin{align} \tag{super-Eddington}  \label{eqn:Lcond}
	\Lw &> \Ledd;\\ \tag{leakage} \label{eqn:tcond}
	\tl &< \tnu, \tcc;\\ \tag{outflow} \label{eqn:outflowcond}
	\tsound,\, \theat,\, \teddy &< \tcc.
\end{align}
Since the spectrum of wave excitation declines  above $\sim 3\, \omc$ \citep{lecoanet2013}, we focus on examining the above conditions for waves with $\omega = 3\, \omc$ throughout each burning phase. For each of the core burning phases, the characteristic $\ell \sim \rprop / H \lesssim 2$, so we focus on waves with $\ell \sim 1$ since these plausibly carry much of the excited wave luminosity and are the most likely to satisfy the \ref{eqn:tcond} condition (see eqn.~\ref{eqn:tleak}). 

For the majority of the shell burning phases, the characteristic $\ell \gtrsim 3$. Assuming the g-mode propagation cavity is determined by the radii $\rprop$ and $\rin$ where $\omega = N$, the leakage timescale (see eqn.~\ref{eqn:tleak}) varies as $\tl \sim \ell^{2 \ell + 1}$ for fixed frequency, since $\rout$ for a wave with degree $\ell$ is given by $S_{\ell}(\rout) = \omega$. Given this  increase in $\tl$ with $\ell$, we conclude that waves excited by shell convection are less likely to satisfy the \ref{eqn:tcond} condition. Thus, we do not specifically address the potential for wave-driven mass loss from shell convection phases, but do provide the total wave energy reservoirs excited by these convective shells in Figs. \ref{fig:oneEwave} and \ref{fig:siEwave}.

Each of the following subsections presents the results of our calculations as follows. We begin by considering each of the \ref{eqn:Lcond}, \ref{eqn:tcond}, and \ref{eqn:outflowcond} conditions in turn. If the most efficiently excited waves in a given progenitor fail either of the first two conditions, there is unlikely to be any significant energy transported from the core to the stellar envelope (i.e., the excited wave energy damps to neutrinos in the core). However, if the progenitor hosts waves that satisfy the \ref{eqn:Lcond} and \ref{eqn:tcond} conditions but fail the \ref{eqn:outflowcond} conditions, there is still a possibility that the energy carried out to the envelope could affect the envelope structure prior to core collapse by e.g., inflating or partially unbinding it.  When considering the \ref{eqn:outflowcond} conditions, there is some uncertainty in whether energy damps radiatively or in the convection zone in some cases. We believe that due to uncertainties in the mixing scheme (i.e., the convective criterion, semiconvection and overshoot), the exact location of the convective boundaries are uncertain at the local scale height level. Thus, for models where $\rdamp$ is within one scale height interior to $\recz$, we consider the waves to be damping within the convection zone.  


\begin{figure*}
   \centering
   \includegraphics[width=0.9\textwidth]{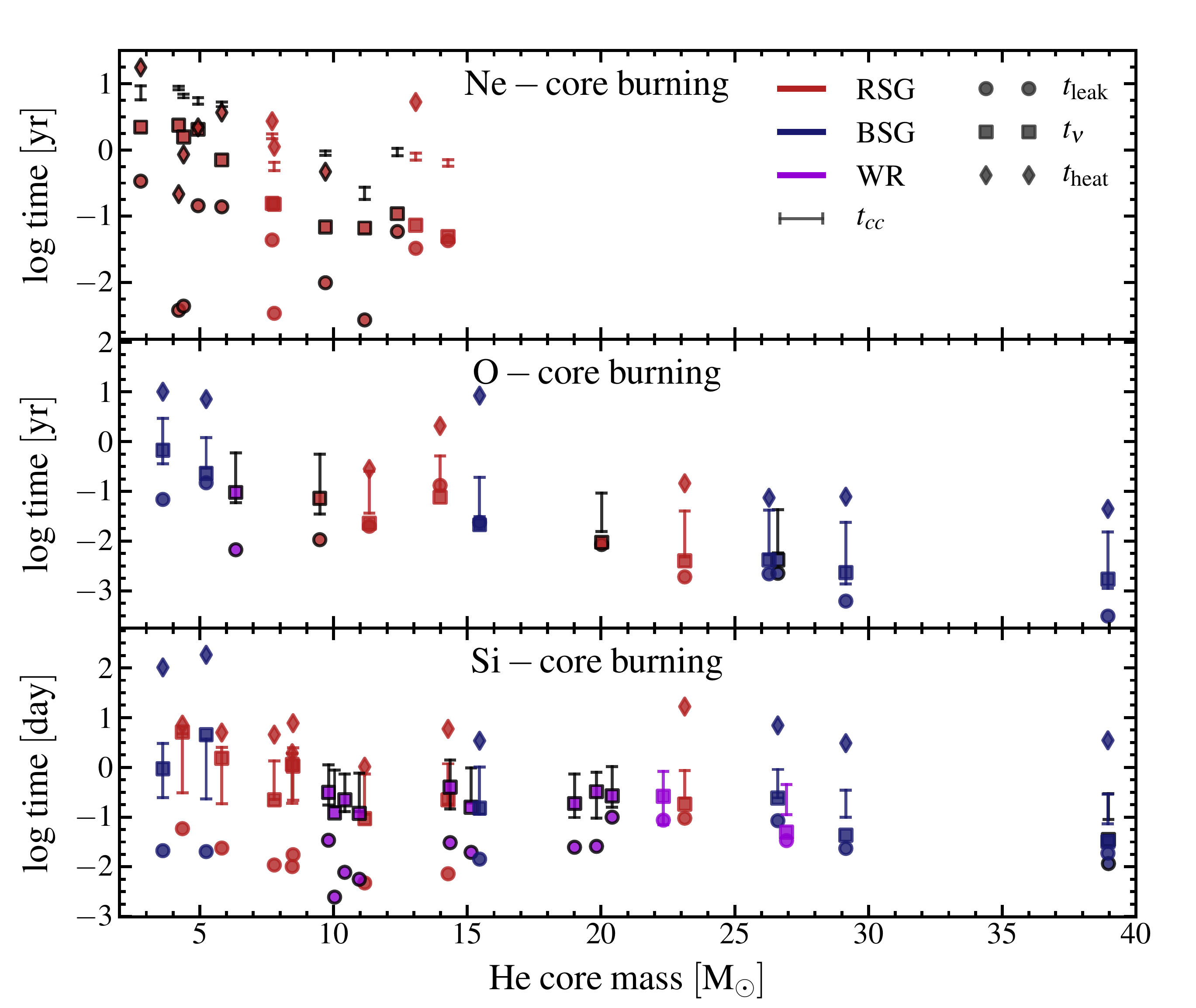}
   \caption{Key timescales for wave driven mass loss calculations during core neon ({\em top}), oxygen, ({\em middle}) and silicon ({\em bottom}) burning, as a function of He core mass at core-collapse.  For clarity, only a randomly chosen 1/3 of our stellar models are shown.   The duration of the burning phase in time to core collapse ($\tcc$) is shown as the capped vertical lines; the timescale for wave energy to leak from the core to the envelope ($\tl$) as filled circles; the core neutrino damping time of the excited waves ($\tnu$) as filled squares; and the timescale to heat the stellar envelope, driving a secondary convection zone ($\theat$), as filled diamonds (when the waves damp in a pre-existing envelope convection zone, the energy is transported outwards on the much faster eddy turnover time [eq. \ref{eqn:teddy}]; $\theat$ is not shown in those cases). Colors represent progenitor classifications as in \S \ref{sec:mods}. Models in which at least 50 per cent of the wave energy satisfies all conditions for a wave-driven outflow ($\tl, \theat, \tnu \lesssim \tcc$) during the given burning phase have points outlined in black.   These are the models most likely to generate significant outflows prior to core collapse.   All timescales  are wave-energy-weighted averages over the duration of the burning phase.} 
   \label{fig:times}
\end{figure*}


To quantitatively illustrate the timescale competition associated with the \ref{eqn:tcond} and \ref{eqn:outflowcond} conditions, Figure \ref{fig:times} shows $\tl$, $\tnu$,  $\theat$, and $\tcc$ as a function of He core mass for neon ({\em top}), oxygen ({\em middle}) and silicon ({\em bottom}) fusion (these are the three burning phases of most interest, as we show in the remainder of this section).   For clarity, we show only a random 1/3 of our stellar models with core convection in the given burning phase.   
All timescales plotted are wave-energy-weighted averages over the duration of the burning phase.\footnote{In the average we exclude points  (i.e., times) that deviate from the median by more than roughly 3 times the dispersion about the median.   This excludes outliers that otherwise strongly bias the average, particularly during silicon burning where there is more fluctuation in the stellar structure that leads to significant temporal variation in $\tl$.} For now, the key point of Figure \ref{fig:times} is that in most models, many of the key timescales that determine the efficacy of wave-driven mass loss are comparable to each other to within a factor of $\sim 10$. Success or failure thus hinges on modest changes in core and envelope structure that change the relevant timescales by factors of $\sim 3-10$. We discuss these in more detail in the following sections.

For progenitors where at least the \ref{eqn:Lcond} and \ref{eqn:tcond} conditions are met, we examine the total wave energy reservoir excited during each phase. This is given by the integral of the wave luminosity over the phase of interest:
\be \label{eqn:eint}
   \Ew = \int \Lw \, dt.
\ee
While we use the above form for our calculations, it is useful to look at an approximate form of eqn.~\ref{eqn:eint} that highlights the important contributions:
\be \label{eqn:ewave}
   \Ew \sim {\rm E_{nuc}} \left( \frac{\Lc}{\rm L_{nuc}}\right) 
      \M^{5/8},
\ee
where ${\rm L_{nuc}}$ is the nuclear luminosity for a given burning phase and ${\rm E_{nuc}}$ the total energy released through fusion during the  phase. 

We regard the determination of the wave energy reservoir as the most definitive conclusion of our work. Our arguments regarding mass loss rates and ejecta masses rely heavily on assumptions about how outflows develop in super-Eddington atmospheres (see \S\ref{ssec:acoustic}), which are as yet not satisfactorily understood. 

Finally, for progenitors where all conditions for wave-driven mass loss are met, we estimate the potential unbound mass, $\Mej$, using eqn.~\ref{eqn:Mej}. This represents an upper limit in the sense that we assume that the outflow taps into the full reservoir of wave energy to produce an outflow (it is not strictly an upper limit because of uncertainties in calculating e.g., $\Ew$ and $\vesc$). We also calculate the radius, $\Rej$, unbound material can reach by core collapse, given simply by $\Rej \approx \vesc \, \tcc$.

Our calculations of the relevant timescales for tunneling and sound-crossing in the envelope rely on accurate determinations of the \brunt{} frequency to determine the relevant radii ($\rprop, \rin, \rout$; see Fig.~\ref{fig:radii}) and the group travel time for the excited g-modes. As the \brunt{} frequency requires numerical differentiation and depends on the implementation of mixing in the stellar interior (in that composition gradients enter into the \brunt{} frequency), our results are affected by numerical noise. Over the course of each burning phase, we evaluate the timescales given in the \ref{eqn:tcond} and \ref{eqn:outflowcond} conditions for at least $\sim 100$ time-steps for each progenitor. We consider any progenitor in which at least 50 per cent of the wave energy excited during a given phase satisfies the \ref{eqn:Lcond}, \ref{eqn:tcond}, and \ref{eqn:outflowcond} conditions as capable of generating wave-driven mass loss throughout that phase. On the other hand, progenitors that fail this 50 per cent criterion are considered incapable of wave-driven mass loss for the given phase. 

In the following, we organize our results by helium core mass at core collapse (rather than ZAMS mass), as this is the best indicator for the time and energy-scales of the late burning phases \citep[e.g.,][]{woosley2002}. 

A summary of all of our wave-driven mass loss calculations is given in Table~\ref{tab:res} in Appendix B. We present a detailed discussion of each of the advanced burning phases in the sections that follow.

\subsection{Carbon-burning} \label{ssec:carbon}


\begin{figure*}
   \centering
   \includegraphics[width=\textwidth]{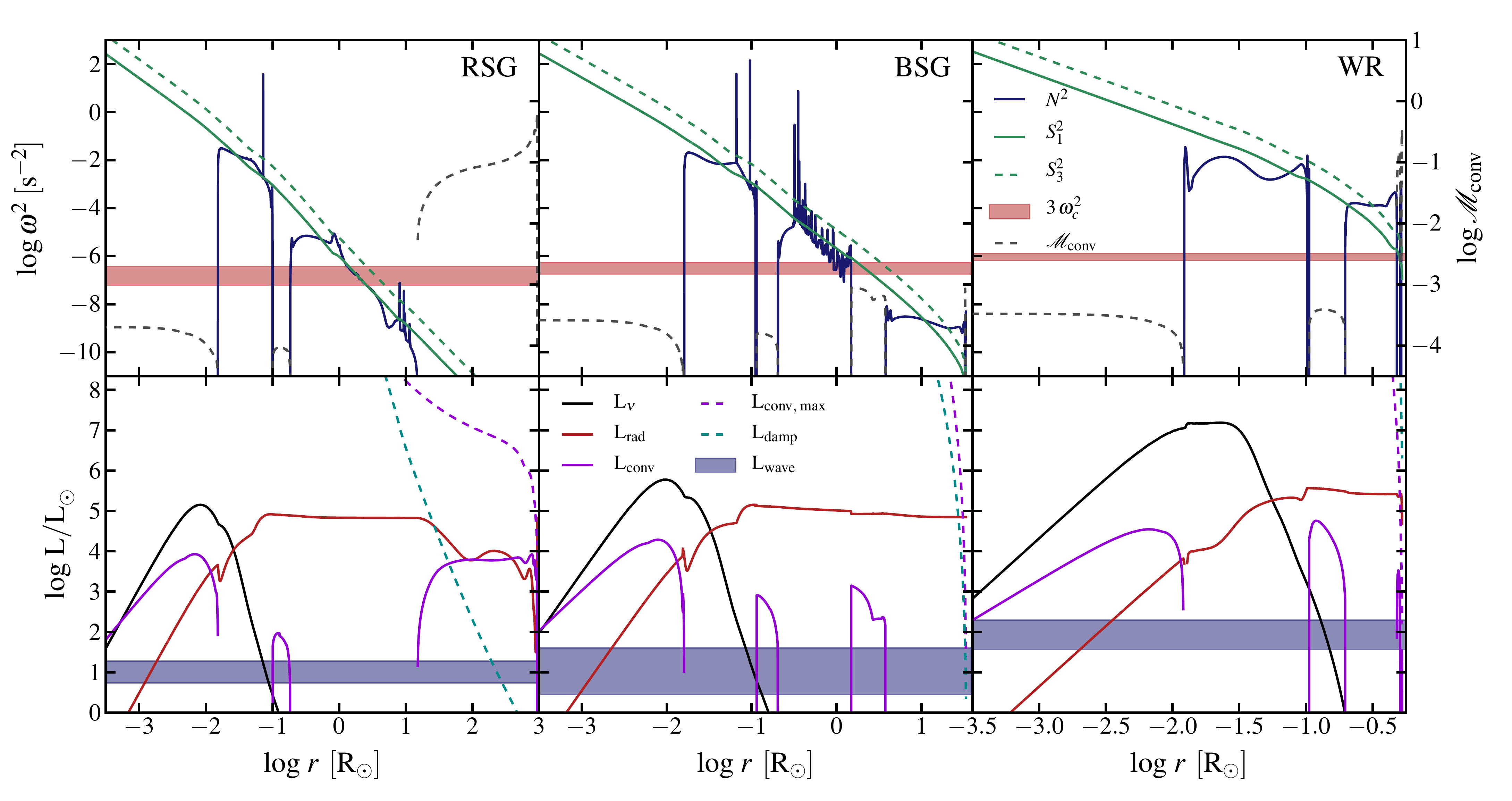}
   \caption{Propagation diagrams and luminosity plots for example core carbon burning models from each progenitor class. Top panels show propagation diagrams, where the colors and dashes are as described in Fig.~\ref{fig:radii} but with the range of propagating g-mode frequencies excited during the whole burning phase shown by the red span. The lower panels show the relevant local luminosities for wave excitation and damping as in the bottom panel of Fig.~\ref{fig:radii} but with the range of wave luminosities excited throughout the burning phase shown as the blue span. These lower panels demonstrate that $\Lw < L_{\rm rad} \sim \Ledd$ during core carbon burning. The red supergiant shown is a 12 \Msun{} solar metallicity model, the blue supergiant a 15 \Msun{} zero metallicity model, and the Wolf-Rayet a 80 \Msun{} solar metallicity model with high mass loss.} 
   \label{fig:cburnPDLum}
\end{figure*} 


Carbon burning is the least interesting of the advanced burning phases from the perspective of wave excitation and subsequent wave-driven mass loss. The nuclear luminosity is generally only slightly larger than the Eddington luminosity and the characteristic mach numbers are the smallest of the post-He burning phases. Furthermore, only progenitors that produce the smallest helium cores (which includes some of the high ZAMS mass progenitors simulated with high mass loss rates) experience convective core carbon burning. The exact helium core mass cutoff depends on the physics implemented in the stellar evolution code, especially the choice of $^{12}{\rm C(\alpha, \gamma)^{16}O}$ reaction rate and the mixing parameters. For large helium core masses, the carbon abundance after helium burning is not large enough for the carbon burning luminosity to exceed the neutrino losses in the core, and carbon is burned radiatively \citep[see e.g.,][]{woosley2002}. 

For our \MESAstar{} calculations, employing the \citet{kunz2002} $^{12}{\rm C(\alpha, \gamma)^{16}O}$ rate, convective core carbon burning occurs only for He-core masses below $\sim 9 \msun$ for non-rotating progenitors. A few of the rotating progenitors with larger helium cores experience convective core carbon burning due to the combination of rotationally enhanced mixing and mass loss. Independent of mass, carbon does burn convectively in a shell after being exhausted in the core; lower core mass progenitors have multiple distinct shell burning phases separated in time (see e.g., Fig.~\ref{fig:kipp12}). 

While core carbon burning excites $\lesssim 10^{46} \erg$ of wave energy over the course of the burning phase, none of the progenitors produce a super-Eddington wave luminosity during this phase of evolution. Fig.~\ref{fig:cburnPDLum} shows propagation diagrams and  plots of the key luminosities for three core carbon-burning models, one of each of our progenitor classes. The top panel shows the \brunt{} frequency (solid blue line), Lamb frequencies for $\ell = 1$ and $3$ (green, solid and dashed lines, respectively) and the range of propagating g-mode frequencies (given by $\sim 3 \, \omc$) predominantly excited during the core carbon-burning phase (red, shaded span) plotted against the left axis; the convective mach number is plotted against the right axis as the dashed grey line. A range of g-mode frequencies is shown to capture variation during the burning phase. 

The bottom panels show the local luminosities important in the physics of wave excitation and damping: the radiative luminosity ($\Lrad$) is shown as the solid red line; the neutrino luminosity ($\Lnu$) in solid black; convective luminosity ($\Lc$) in solid magenta; the maximum possible convective luminosity ($\Lcmax$) in dashed magenta; the damping luminosity for radiation ($\Ldamp$) in dashed teal; and the range of $\Lw$ excited during this phase as the blue, shaded span. All progenitors that undergo convective core carbon burning are qualitatively similar, in that $\Lw < \Ledd$ throughout the phase. Thus, wave-driven mass loss is unlikely to arise from core carbon burning.

\subsection{Neon burning} \label{ssec:neburn}

As for core carbon burning, only lower mass models (helium core masses $\lesssim 16 \msun$ in this case) have a distinct convective core neon burning phase. This is shown in Fig.~\ref{fig:kipp12} as the short core convection phase just after the growth of the carbon depleted core ($\log \tcc \sim 0.75$) and similarly in Fig.~\ref{fig:kipp40} ($\log \tcc \sim 0$). Also similar to core carbon burning, the mass fraction in Ne left behind by earlier phases decreases with core mass, so that the net nuclear luminosity, and thus the wave luminosity, decreases with increasing core mass. Thus, only 43 of our 76 progenitors have distinct core neon burning phases.


\begin{figure*}
   \centering
   \includegraphics[width=\textwidth]{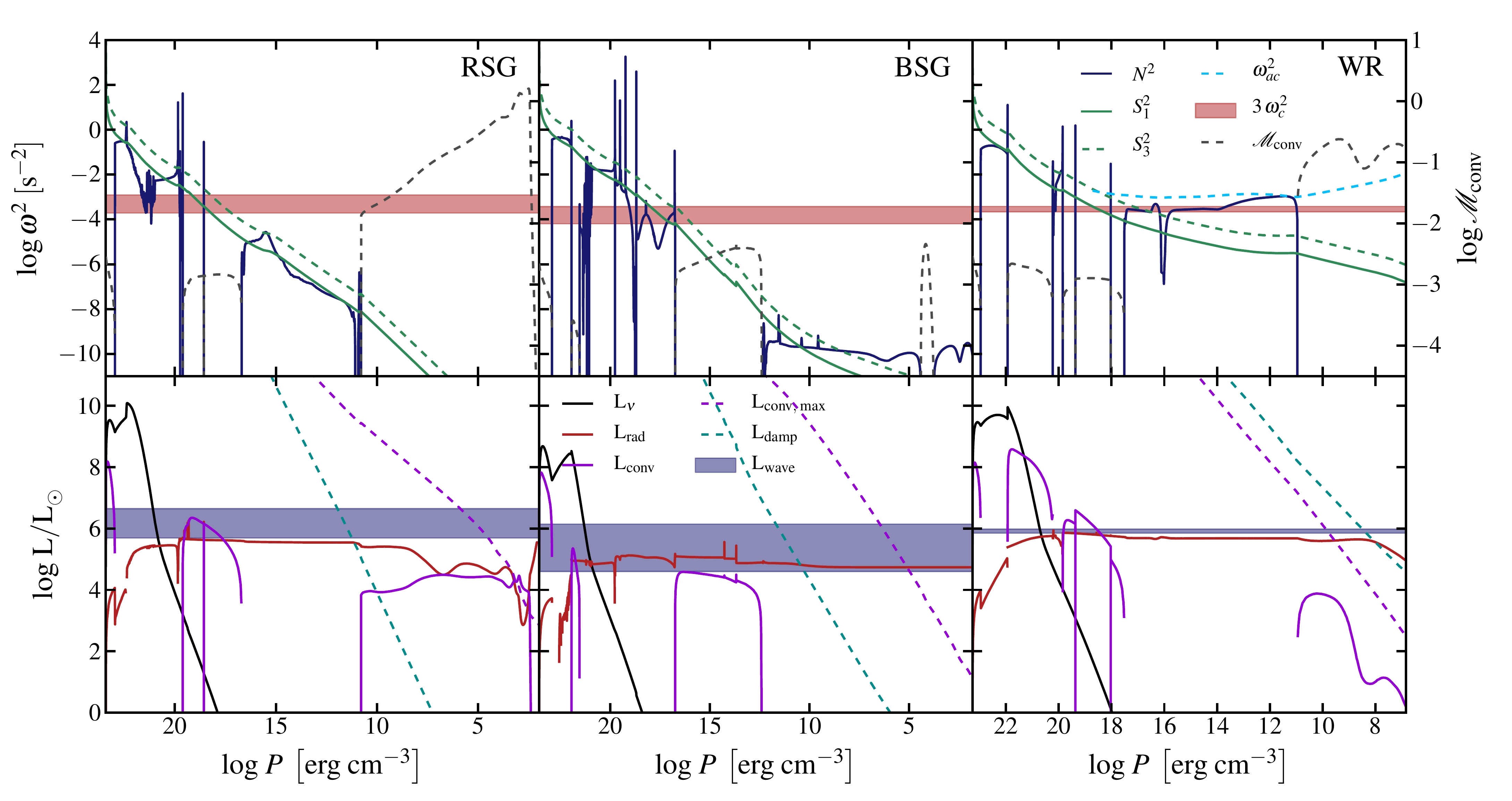}
   \caption{Propagation diagrams and luminosity plots for example core neon burning models from each progenitor class.   The abscissa is log[Pressure] to highlight the surface behavior.   The colors and dashed lines are as described in Fig.~\ref{fig:cburnPDLum}, with the addition of the acoustic cutoff frequency ($\omac$) in the envelope shown as the dashed cyan line in the top right panel for the Wolf-Rayet progenitor. This is plotted to show that the g-mode frequencies are below the acoustic cutoff for this representative compact progenitor, and thus waves are likely reflected before reaching the envelope. For the red/blue supergiant progenitors, $\omac$ roughly follows the Lamb frequency and is never larger than the g-mode frequencies so waves easily propagate into the stellar envelope. The red supergiant shown is a 25 \Msun{} solar metallicity model, the blue supergiant a 15 \Msun{} zero metallicity model, and the Wolf-Rayet a 40 \Msun{} solar metallicity model.} 
   \label{fig:neburnPDLum}
\end{figure*}


All neon burning RSG and BSG progenitors excite gravity waves that meet the \ref{eqn:tcond} condition, but only models with helium core masses $\lesssim 14 \msun$ generate a super-Eddington wave luminosity during core neon burning. For the compact WR progenitors, core-Ne burning predominantly excites waves with frequencies $\lesssim \omac$ in the stellar envelope, which are thus unlikely to tunnel out of the stellar core. This leaves 27 of the 43 core neon burning progenitors where waves are likely to transport energy out of the core and into the stellar envelope.

Figure~\ref{fig:neburnPDLum} shows the propagation diagrams and luminosity plots as in Fig.~\ref{fig:cburnPDLum}, but with $\log P$ on the abscissa rather than $\log r$, in order to show the envelope behavior. The top right panel showing the propagation diagram for the WR progenitor also shows the acoustic cutoff frequency, $\omac$, in the envelope (dashed, cyan line) to demonstrate its magnitude in comparison to the excited g-mode frequencies. That $\omega \lesssim \omac$ is also reflected in the bottom right panel where $\Ldamp < \Lcmax$ at all radii; this is equivalent to $kH < 1$ (see eqns.~\ref{eqn:ldamp} and \ref{eqn:lbreak}). This representative example demonstrates that convectively excited waves in WR progenitors during core neon burning are likely reflected before reaching the envelope. For the giant progenitors, $\omac \ll \omega$ for the excited g-modes and roughly follows the Lamb frequency (so it is not shown). In giants that satisfy the \ref{eqn:Lcond} and \ref{eqn:tcond} conditions, waves likely tunnel out of the core and deposit their energy at $\rdamp$.

\begin{figure}
   \centering
   \includegraphics[width=\columnwidth]{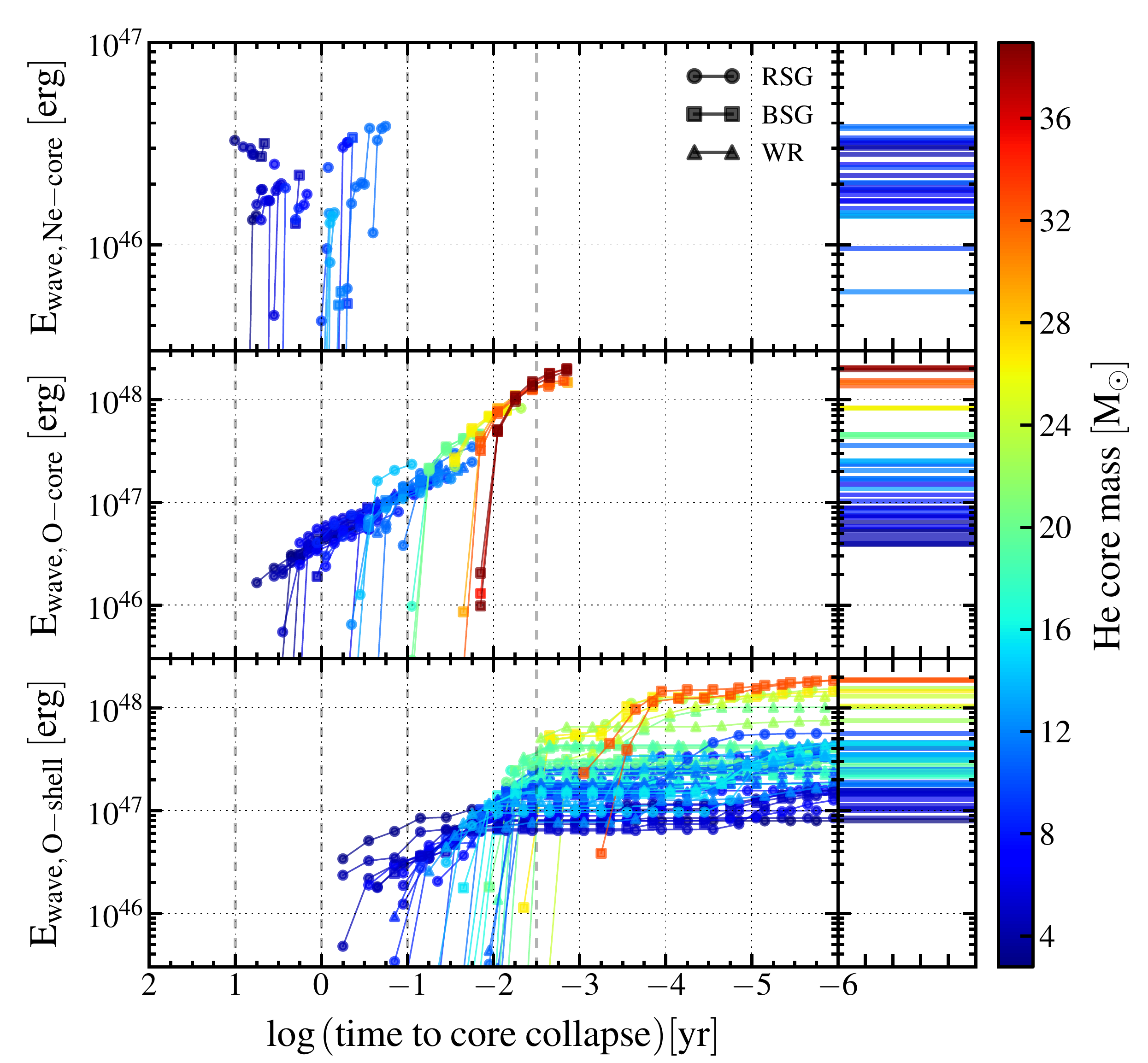}
   \caption{Wave energy excited (see eqn.~\ref{eqn:ewave}) during core-Ne and O burning ({\em top and middle left} panels, respectively) and shell-O burning ({\em bottom left} panel) for each of the progenitors that satisfies the \ref{eqn:Lcond} and \ref{eqn:tcond} conditions, plotted against time to core collapse $\tcc$. Colors correspond to helium core mass, as given by the color bar at right. The integrated energy, from the beginning of the burning phase up to a given $\tcc$, is plotted every $0.05$ dex in $\log \tcc$ in the top panel, $0.2$ dex in the middle panel, and $0.3$ dex in the bottom panel. {\em Top and middle right} panels give the total (cumulative) wave energy excited over the entirety of each core burning phase. {\em Bottom right} gives the total energy excited over the shell burning phase (or the total at $\log \tcc \sim -6$, whichever occurs first). The shape of the points correspond to progenitor type as noted in the legend in the top panel. These plots demonstrate the near one-to-one correspondence between helium core mass, available wave energy, and time to core collapse (see \S \ref{ssec:oburn}).}
   \label{fig:oneEwave}
\end{figure}

The upper panels of Fig.~\ref{fig:oneEwave} show the total energy released in waves during core neon burning for all progenitors that satisfy the \ref{eqn:Lcond} and \ref{eqn:tcond} conditions. In the upper left panel, the integral in eqn.~\ref{eqn:eint}, evaluated from the start of the burning phase to a given $\tcc$, is plotted every $\log \tcc \sim 0.05$. The total (cumulative) energy liberated over the whole burning phase for each model is shown as a horizontal line in the upper right panel. The colors represent the helium core mass, according to the colorbar at the far right of the plot, and the symbols represent the progenitor type, with circles representing RSGs, squares BSGs, and triangles WRs. There are no strong trends in $\Ew$ during core neon burning.

For all progenitors that satisfy the \ref{eqn:Lcond} and \ref{eqn:tcond} conditions (giants), waves reach $\rdamp$ interior to $\rss$, on a timescale $\tsound \sim 10^{4} \s \ll \tcc$ (since $\rdamp \ll R_\star$). For four of the RSGs, $\rdamp$ is within one scale height of the envelope convection zone; we consider the timescale for driving an outflow from these few progenitors to be $\teddy \ll \tcc$. However, the majority, 23, reach $\rdamp$ in the radiative zone $\sim 10$ scale heights from the outer convection zone and must heat the stellar material to drive convection and eventually an outflow. Calculating $\theat$ over the region from $\rdamp$ to the base of the envelope convection zone $\recz$, we find 14 of the remaining 23 models have $\theat < \tcc$, leading to the conclusion that wave-driven mass loss is plausible in these 14 progenitors. The remaining nine stars would reach collapse with $\sim 10^{46} \erg$ of wave energy still attempting to work its way through the envelope.


The upper panel of Fig.~\ref{fig:oneMej} shows our estimate of $\Mej$ (see eqn.~\ref{eqn:Mej}) for the 18 progenitors (out of 43 convective core-Ne burning and 76 total progenitors) where wave-driven mass loss is possible during core neon burning. Our RSG progenitors, having the most weakly bound envelopes, are capable of producing the largest wave-driven mass loss events, with $\Mej \sim 0.1 - 1 \msun$. The two core-Ne burning BSGs capable of driving outflows can produce $\Mej \lesssim 0.04 \msun$. If launched during core-Ne fusion, traveling at the escape velocity at $\rss$, $\sim 100{\rm s} \kms$, this ejecta can reach distances of $\lesssim 300 \AU$ prior to core collapse. Lower mass progenitors, with their smaller helium cores, longer burning timescales, and tendency to form giants, are capable of producing the most massive and extended wave-driven circumstellar environments during Ne fusion.

\subsection{Oxygen burning} \label{ssec:oburn}

Core oxygen burning is convective for all progenitors, but occurs over a range of timescales \citep[see Table~\ref{tab:mods}, and][]{woosley2002}; the burning timescale depends primarily on helium core mass, with the smallest cores ($\sim 3 \msun$) having core oxygen burning for $\sim 6 \yr$ and the largest ($\sim 40 \msun$) for $\sim 10 \,{\rm day}$. For all the progenitors considered here, the burning produces a super-Eddington wave luminosity during the majority of  core oxygen burning.

In all but five of the 42 RSG and BSG progenitors, convectively excited waves can transport energy from the core to the envelope. As expected from eqn.~\ref{eqn:tleak}, those that fail are biased towards larger values of $\rout / \rin$. For the WRs, only three out of 34 compact progenitors have energy-bearing waves with frequencies above the envelope acoustic cutoff during core oxygen burning. This leaves a total of 40 (37 giants and 3 WRs), a little more than half of the progenitors, capable of transporting wave energy from the core to the envelope during core oxygen burning. 


\begin{figure}
   \centering
   \includegraphics[width=\columnwidth]{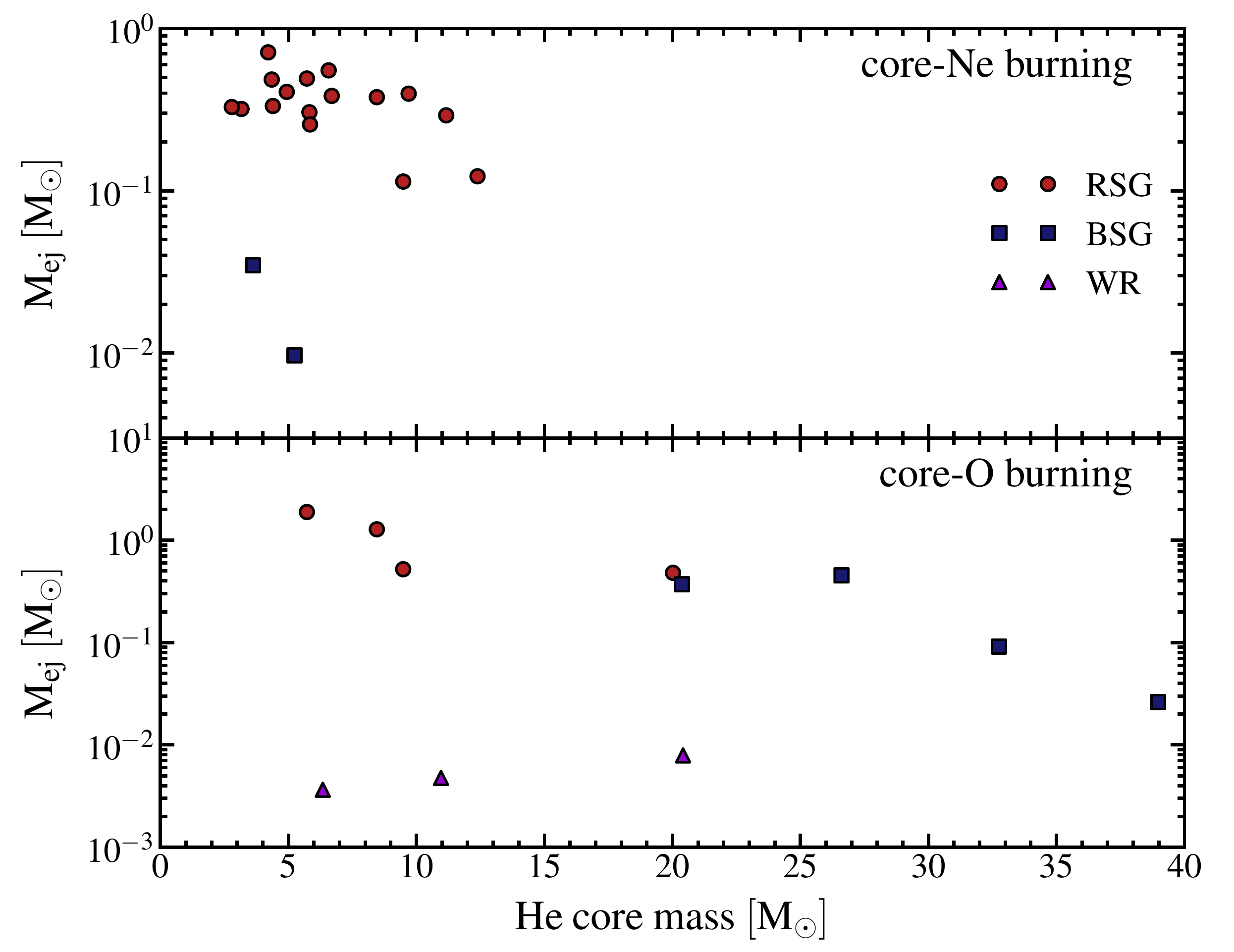}
   \caption{Potential wave-driven ejecta mass generated during core neon ({\em top} panel) and oxygen burning ({\em bottom} panel), estimated using eqn.~\ref{eqn:Mej}. The colors and symbols represent progenitor types as noted in the legend. Note the clear delineation of $\Mej$ between giants and Wolf-Rayets, due to differences in the envelope binding energy.}
   \label{fig:oneMej}
\end{figure}


The middle panel of Fig.~\ref{fig:oneEwave} shows the wave energy reservoir during core oxygen burning for these 40 progenitors, while the bottom panel shows the wave energy from shell oxygen burning. There is a factor of 50 increase in the wave energy reservoir from core oxygen burning as the helium core mass increases from $\sim 3.2 - 39 \msun$. This increase arises partially from each of the contributions outlined in eqn.~\ref{eqn:ewave}. As core mass increases, the time averaged mach number during the oxygen core burning increases by a factor of $\sim 20$, the total nuclear energy liberated varies by a factor of $\sim 5$, and the ratio of the convective to nuclear luminosities varies by a factor of $\sim 2$. While the ratio $\Lc / \Lnuc$ varies only slightly, the ratio $\Lw / \Ledd$ varies from $\sim 1 - 10^3$ due to the change in burning timescale and thus $\Lnuc$. 

During the short core oxygen burning phase, fewer progenitors host waves capable of heating the stellar envelope on a timescale shorter than the time to core collapse. Only one of the 40 progenitors with waves capable of transporting a super-Eddington luminosity to the envelope has $\theat < \tcc$ and can drive an outflow via radiatively damped waves. Four others excite waves that reach $\rss < \rdamp$, but have $kH > 1$, so that they can drive an outflow on $\sim$ the sound crossing time at $\rss$, which is characteristically on the order of minutes. Finally, six giants have waves with $\rdamp$ less than one scale height from the envelope convection zone; for these, we assume the deposited wave energy can immediately drive convection into the envelope zone and drive an outflow on a timescale $\teddy \lesssim \, {\rm day} < \tcc$. In total, 11 of our 76 progenitors are capable of generating wave-driven mass loss during the core oxygen burning phase; these 11 span all three progenitor categories, as well as our full range of initial masses, metallicities and rotation rates. 

The lower panel of Fig.~\ref{fig:oneMej} shows our estimate of $\Mej$ for the 11 progenitors capable of wave-driven mass loss during core oxygen burning. As during core-Ne burning, the weakly bound envelopes of the giant progenitors are most susceptible to large mass-loss events. During core-O burning, RSGs may produce $\Mej \sim 0.3 - 3 \msun$, BSGs may liberate $\Mej \lesssim 1 \msun$, and WRs $\lesssim 0.01 \msun$. Given the similar timescales and assumed envelope structures for oxygen and neon fusion, the outflow velocities and the radii reached by the ejecta during core-O burning are similar to those during Ne burning, with ejecta reaching $\lesssim 300 \AU$ at speeds of $\gtrsim 100{\rm s} \kms$. 

\subsection{Silicon burning} \label{ssec:siburn}

The silicon burning phase is the most uncertain of those considered here. During this late stage, burning and convective timescales become comparable, likely invalidating the treatment used in most stellar evolution codes, including \MESAstar{} \citep[e.g.,][]{woosley2002}. Furthermore, the reaction network uses many pseudo-reactions to simulate the actual high-dimensional nucleosynthetic network (though development is underway to improve upon this; Paxton, \textit{private communication}). These uncertainties may affect the luminosities and stellar structure during silicon burning.  In addition, we find that the wave leakage timescale $\tl$ (eq. \ref{eqn:tleak}) is much more variable from timestep to timestep during silicon burning, due in part to the existence of multiple non-overlapping convection zones that cause large fluctuations in the \brunt{} frequency.

In our \MESAstar{} models, silicon burns convectively in the core for all progenitors, over timescales of $\lesssim 5 \, {\rm day}$ for the smallest cores down to $\sim 6 \hr$ for the largest. The most luminous burning phase, core silicon burning produces a significantly super-Eddington wave luminosity in all progenitors, $\Lw \sim 10^2 - 10^4 \, \Ledd$. The characteristic excitation frequency, $\sim 3\, \omc$, increases from $\sim 10^{-3} \Hz$ during carbon burning to $\sim 0.1 \Hz$ during this last phase. At these high frequencies, convectively excited waves now exceed the acoustic cutoff frequency in all compact progenitors.   This is a key difference between silicon fusion and earlier burning phases.


\begin{figure}
   \centering
   \includegraphics[width=\columnwidth]{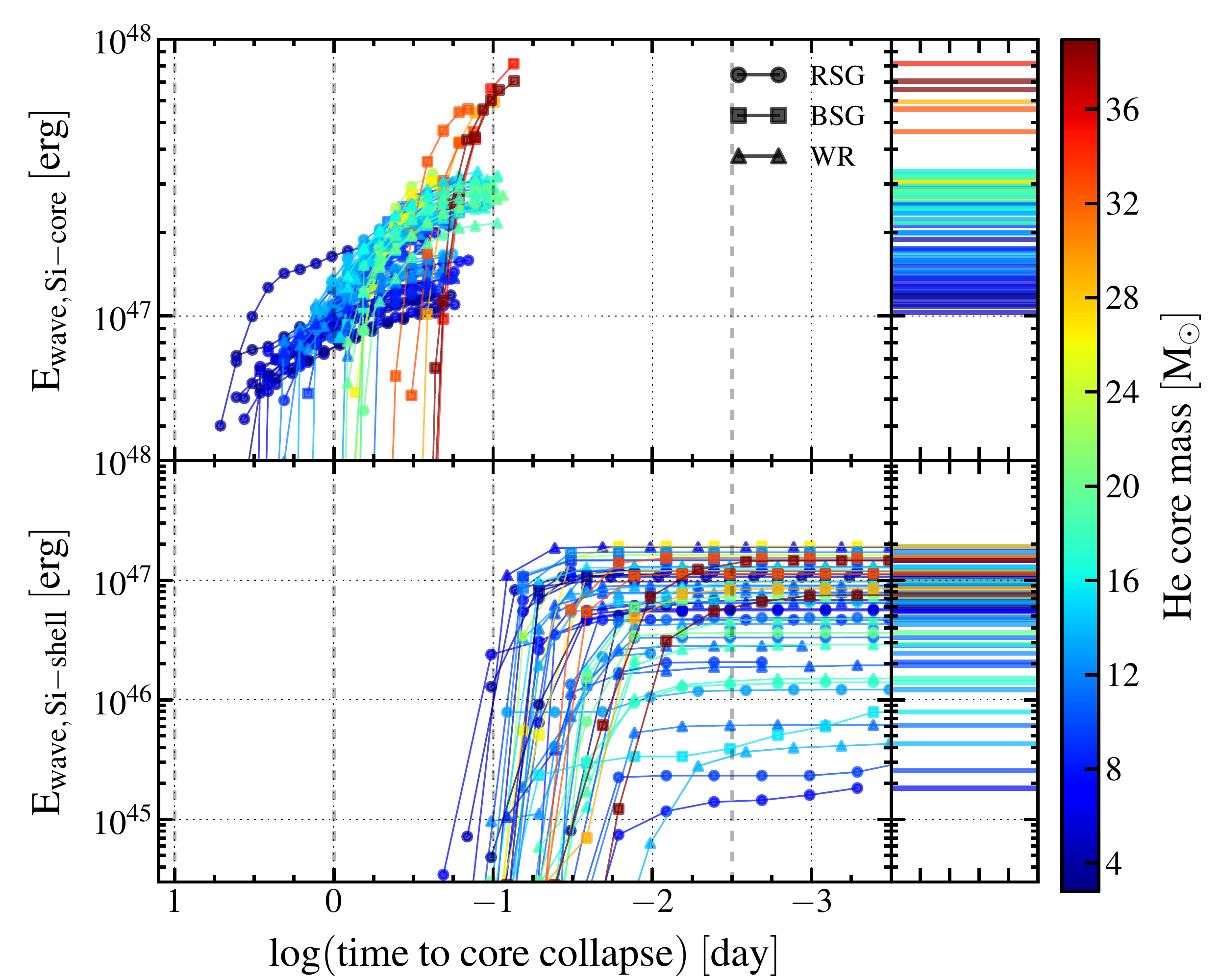}
   \caption{Wave energy excited (see eqn.~\ref{eqn:ewave}) during core ({\em top left} panel) and shell silicon burning ({\em bottom left} panel) for each of the progenitors that satisfies the \ref{eqn:Lcond} and \ref{eqn:tcond} conditions, plotted against time to core collapse \textit{in days}. {\em Top right} panel gives the total (cumulative) wave energy excited over the whole core-Si burning phase; {\em bottom right} panel shows the total energy excited during shell-Si burning (or the total at $\log (\tcc/{\rm day}) \sim -3.5$, whichever occurs first).  Colors correspond to helium core mass, as given by the color bar at right.   The shape of the points correspond to progenitor type as noted in the legend in the top panel.}
   \label{fig:siEwave}
\end{figure}


With the compact progenitors exciting waves above the acoustic cutoff frequency, many more of our progenitors transport energy from the core to the surface during core silicon burning than during the previous phases: 70 of 76. Figure~\ref{fig:siEwave} shows the wave energy reservoir for these 70 progenitors during core silicon burning. There is a roughly monotonic relationship between $\Ew$ and helium core mass, as in the case of core oxygen burning, but the range in $\Ew$ is smaller at a factor of $\sim 8$.

Thirty-one of the seventy progenitors that satisfy the \ref{eqn:Lcond} and \ref{eqn:tcond} conditions also satisfy the \ref{eqn:outflowcond} conditions. Of these, the majority, 28, are WR progenitors which now host convectively excited waves with frequencies above the acoustic cutoff, as described above. In these, and one BSG progenitor, waves reach $\rss$ before $\rdamp$, likely generating outflows on the sound crossing time at $\rss$, which is $\lesssim$ minute. Only two RSG progenitors host waves that satisfy the \ref{eqn:outflowcond} conditions; one where $\rdamp$ is within one scale height of the envelope convection zone, and one where $\rdamp$ is in the radiative zone and $\theat, \teddy < \tcc$. Despite the very short time to core-collapse, $\tcc$, the several order of magnitude increase in $\Lw$ during silicon fusion sufficiently decreases $\theat$ for this one progenitor.

\begin{figure} 
   \centering
   \includegraphics[width=\columnwidth]{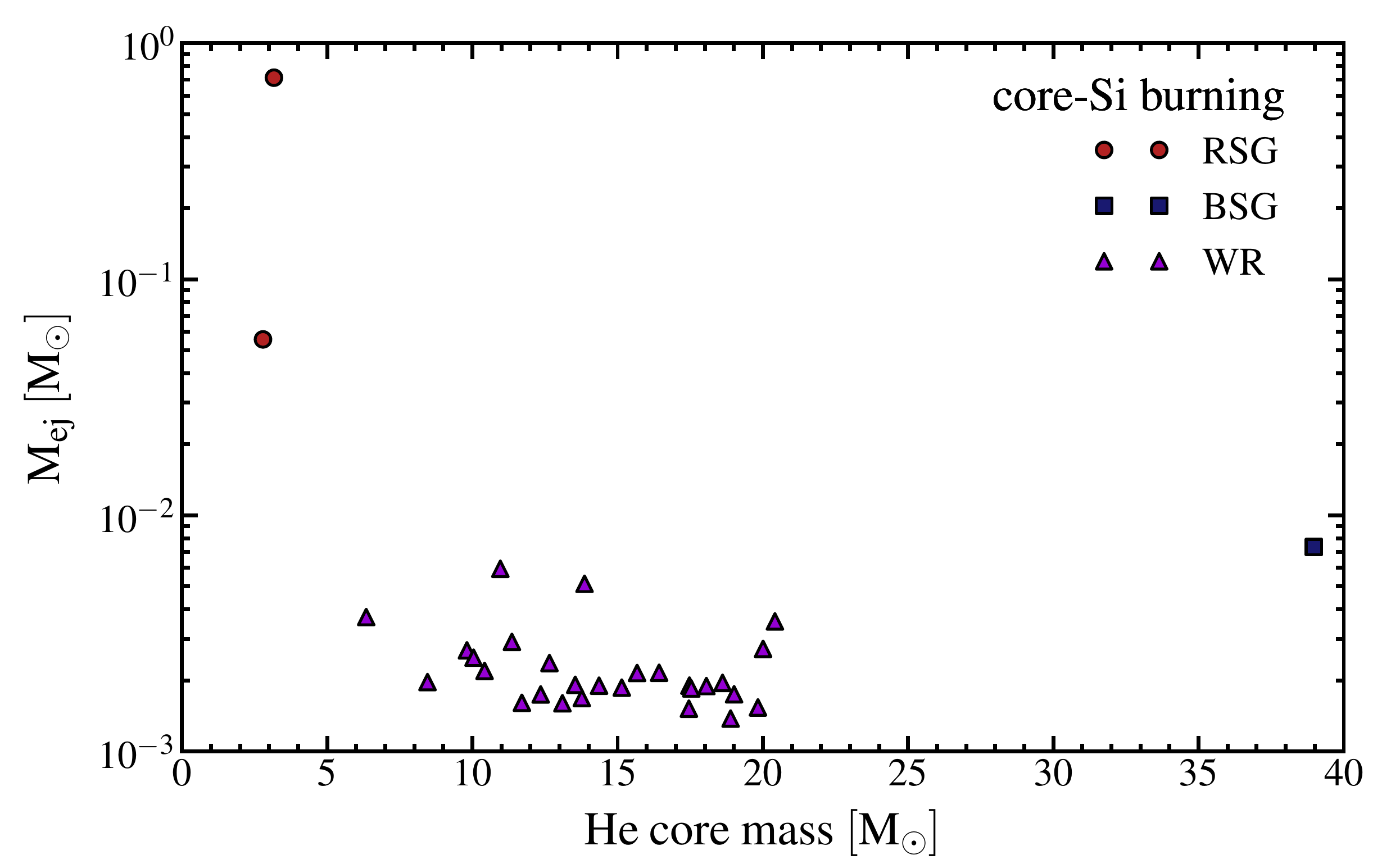}
   \caption{Potential wave-driven ejecta mass generated during core silicon burning.  The legend shows the colors and symbols representing each progenitor type. As in earlier burning phases, there is a clear delineation in $\Mej$ by progenitor type due to differences in the envelope binding energy. For the giant progenitors, there is not enough time before core collapse to eject material beyond the stellar radius. Instead, the wave energy will significantly restructure the stellar envelope prior to explosion.} \label{fig:siMej}
\end{figure}

Figure~\ref{fig:siMej} shows the mass, $\Mej$, that can be unbound by waves excited during core-Si burning. As during previous phases, there is a clear delineation in $\Mej$ between giants and WRs due to differences in the envelope binding energy. While two RSGs are capable of heating the stellar material and potentially driving an outflow prior to core collapse, there is not enough time for any unbound mass to travel beyond the progenitor photosphere, which is already $\gtrsim 100 \rsun$. However, for the WR progenitors, wave energy deposited in the envelope during Si burning has the potential to inflate these compact progenitors to giants extending out to $100{\rm s} \rsun$, in the $\sim$ day leading up to collapse. This may have important implications for the appearance of a subsequent SNe.


\section{Discussion and conclusions} \label{sec:disc}

In \citet{quataert2012}, we argued that vigorous core convection in late stages of stellar evolution (Ne fusion and later) can excite a super-Eddington energy flux in outgoing internal gravity waves, potentially leading to substantially enhanced mass loss in the last year to decade of stellar evolution.   The ultimate energy source for this mass loss is the prodigious fusion luminosities in the cores of massive stars during their neutrino-cooled phases.

In this work, we have surveyed SN progenitors with initial masses from $12 - 100 \msun$, metallicities from 0 (Population III) to solar and rotation up to 80 per cent of breakup in search of those most susceptible to  wave-driven mass loss. This grid of stellar models likely provides a reasonable sample of the range of possible helium core masses and progenitor structures produced during single star evolution.   Binary evolution may lead to qualitatively different progenitors.   Moreover, the mapping from ZAMS properties to late-stage stellar properties is uncertain because it depends on  details of mass and angular momentum loss throughout stellar evolution.

For many progenitors, we find that waves excited  after core carbon exhaustion can carry a super-Eddington luminosity and deposit $10^{46} - 10^{48} \erg$ out in the stellar envelope.   Depending on the uncertain physics of super-Eddington stellar envelopes, this may lead to strong wave-driven mass loss in about 20 per cent of the progenitors surveyed.    While a detailed comparison to SN rates depends on the initial mass function, the connection between ZAMS and helium core masses, and the subset of massive stars that can explode, we note that the rate of wave-driven mass loss we find is comparable to the rate of Type IIn SNe, which are $\sim 10$ per cent of core collapse SNe \citep{li2011}.    

There are several physical properties of the stellar progenitor that determine whether or not wave-driven mass loss is likely (see \S \ref{sec:theory}).    In those progenitors we have surveyed where wave-driven mass loss appears disfavored, the most common reasons are (1) the wave energy is thermalized by dissipation  deep in the star where the timescale to generate a convection zone that carries energy to the stellar surface ($\theat$; eq.~\ref{eqn:theat}) is longer than the time to core collapse, (2) for compact WR-like progenitors, the core convective frequency during neon and oxygen burning is below the acoustic cutoff frequency of the stellar envelope;  waves excited by core convection thus cannot propagate out into the stellar envelope.


\begin{figure}
   \centering
   \includegraphics[width=\columnwidth]{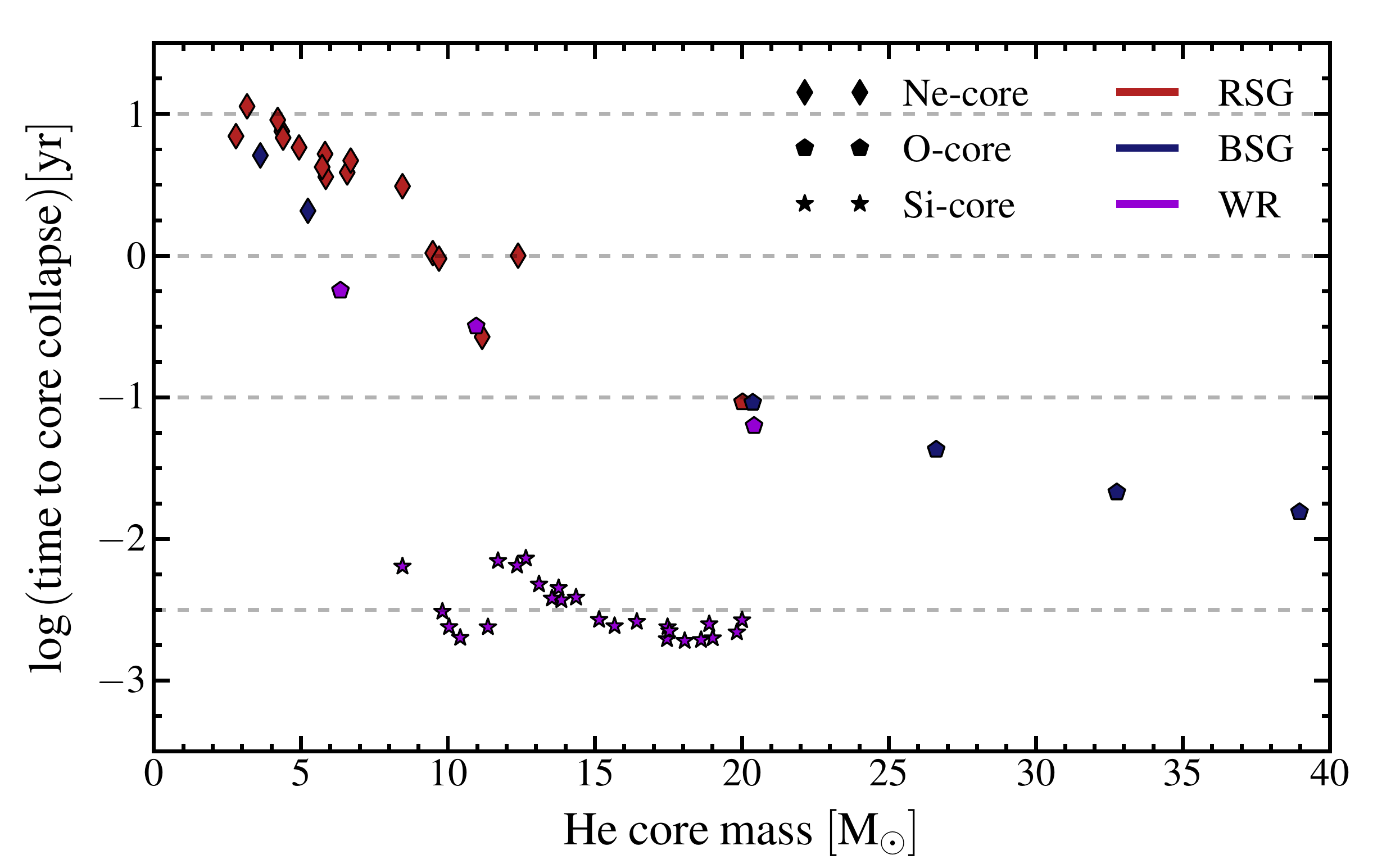}
   \caption{Earliest potential onset of wave-driven mass loss for SN progenitors, plotted as a function of helium core mass. Symbols denote the phase of burning corresponding to the earliest potential onset of wave-driven mass loss. Colors represent progenitor type as in previous figures. Grey dashed lines mark 10 years, 1 year, 1 month, and 1 day prior to core collapse (from top to bottom). Note that lower core masses produce wave-driven mass loss significantly earlier in evolution.} \label{fig:tonset}
\end{figure}


Given that the wave luminosity excited by core carbon fusion is never super-Eddington, we find that the earliest core burning phase that might lead to wave-driven mass loss is neon fusion.  Figure~\ref{fig:tonset} shows the earliest potential onset of wave-driven mass loss ($t_{\rm onset}$) as a function of helium core mass. For the lowest core masses, the onset of super-Eddington wave luminosities may be as early as $\sim 10 \yr$ before core collapse.   Figure \ref{fig:tonset} also highlights the well known \citep[e.g.,][]{woosley2002} correlation between burning duration (and thus $t_{\rm onset}$) and core mass, which is one of the key features of our specific mass loss mechanism. For wave-driven mass loss events, the timing of a pre-SN outburst can be used to place a rough upper limit on the helium core mass of the progenitor, which can in turn be constrained by modeling the associated SN. For example, a burst of mass loss that occurs more than a month prior to core collapse requires a progenitor with a helium core mass $\lesssim 15 \msun$.  However, the complex interplay between mixing and mass loss, each dependent on both rotation and metallicity, makes the further inference from helium core mass to ZAMS mass, metallicity and rotation much less certain.   Even the surface luminosity and effective temperature at a given helium core mass are  uncertain because of difficulties modeling near-Eddington stellar envelopes (e.g., \citealt{mesaII,suarez2013}; see also \S \ref{sec:mods}).

Table~\ref{tab:res} in Appendix B summarizes the results of our calculations. For each model, the time to core collapse ($\tcc$) for each of the neon, oxygen and silicon burning phases is given whenever these phases occur convectively. The wave energy reservoir ($\Ew$) for each burning phase is given for models in which convectively excited waves can tunnel out of the core and carry a super-Eddington luminosity into the envelope. Finally, when this excited wave energy can plausibly drive an outflow on a timescale shorter than $\tcc$, we present the estimated ejecta mass ($\Mej$), escape velocity ($\vesc$) at the point where a wind is plausibly launched (i.e., $\rss$), and the radius that ejecta reaches prior to explosion ($\Rej$). These results are also summarized in Figs.~\ref{fig:oneEwave}---\ref{fig:tonset}.   As noted earlier in the paper, the ejecta energetics $\Ew$ are more robust  than the ejecta mass $\Mej$ or radius $\Rej$ because the former ``only" depends on the physics of wave excitation, propagation and damping while the latter also depend on the physics of outflows under super-Eddington conditions.

We find that lower mass stars are the most likely to produce significant circumstellar ejecta via wave-driven mass loss during neon and oxygen fusion.   In more massive stars, wave-driven mass loss during neon/oxygen fusion is also possible but occurs preferentially at lower metallically, $Z \sim 0.01-0.1 Z_\odot$.   This is because massive solar metallically stars become WR stars prior to core-collapse with our mass-loss prescriptions.   The convectively excited waves during oxygen and neon fusion in most WR progenitors have frequencies below the acoustic cutoff frequency of the stellar envelope, thus suggesting that the convectively excited waves from the core cannot efficiently heat the stellar envelope.  We note, however that the wave frequencies in our models are only a factor of few below the acoustic cutoff frequency (e.g., the top right panel of Fig.~\ref{fig:neburnPDLum}).  In a small minority of our WR progenitors we thus find that wave-driven mass loss during core neon and oxygen fusion may be possible (e.g., the $20 \msun$, $Z = Z_\odot$, and $\Omega = 0.8 \, \Omega_{\rm crit}$ and $40 \msun$, $Z = Z_\odot$ and $\Omega = 0.2 \, \Omega_{\rm crit}$ models in Table \ref{tab:res}).

Progenitors with small helium cores are susceptible to the earliest mass loss events, up to $\sim 10 \yr$ prior to core collapse (Fig. \ref{fig:tonset}).   This is important  because mass ejection in the year to decade prior to core-collapse  places the resulting ejecta at radii $\sim 100$ AU at explosion.  At these radii, the shock produced by the interaction between the SN and the previously ejected matter is particularly radiatively efficient, thus enabling circumstellar interaction to efficiently power luminous SNe.     In all cases, the lower binding energy of the stellar envelope in blue and red super-giants  implies that these progenitors are the most likely to produce very massive ejecta prior to core-collapse.

During silicon burning in the last few days of stellar evolution, there is not enough time for wave energy deposition in super-giant progenitors to significantly alter the stellar radius, though the wave energy may modify the structure of the stellar envelope.   However,  wave excitation and damping during Si burning can inflate nominally compact WR progenitors to giant radii ($\sim 100{\rm s} \rsun$).      The key difference relative to neon and oxygen fusion, during which WR progenitors are less likely to have significant wave-driven mass loss, is that the convection is far more vigorous during silicon fusion.  This excites higher frequency waves that exceed the acoustic cutoff frequency of the envelope.    The end result of wave energy deposition during silicon fusion in WR progenitors would likely be a core-collapse SN spectroscopically classified as Type Ibc (i.e, a compact star) but with early thermal emission more characteristic of extended shock-heated stellar envelopes.   

The success of the neutrino mechanism for core-collapse SNe hinges on a competition between neutrino heating of the post accretion shock material and the ram pressure of the accreting matter (e.g., \citealt{burrows1993}).    Any process that decreases the accretion rate  in the first $\sim$ sec after core bounce can aide the explosion.   The energy in convectively-excited waves that {break} and dissipate just outside the silicon-burning core is almost certainly significantly larger than the energy that makes its way to the stellar envelope.  Moreover, this redistribution of energy occurs for {\em all} stellar progenitors.   It is likely that the dissipated energy modifies the stellar density profile relative to existing pre-SN models:   e.g., at a radius of $\sim 10^9$ cm the deposition of $\sim 10^{49}$ ergs of wave energy during silicon burning could double the radius of $\sim 0.04 \, M_\odot$.   Whether this is sufficient to significantly aide the onset of explosion remains to be seen.   This effect can be crudely taken into account in 1D stellar evolution models or more self-consistently modeled using multi-dimensional hydrodynamic simulations (e.g., \citealt{meakin2006,arnett2011}).

\subsection{Comparison to observed systems} \label{ssec:obs}

Here we discuss the application of our results to Type IIn and Type IIb SNe, the latter of which appear to arise from both compact and extended progenitors \citep{chevalier2010}.   In addition, there are  three SNe with {observed} pre-SN outbursts: the Type Ibn 2006jc \citep{foley2007,pastorello2007}, the Type IIn 2009ip \citep[e.g.,][]{mauerhan2013,fraser2013,margutti2013}, and the Type IIn 2010mc/PTF 10tel \citep{ofek2013}. We briefly address each of these systems here. 

\subsubsection{Type IIn SNe}

Type IIn SNe are characterized by the presence of narrow lines in their spectra, indicative of interaction between the outgoing SN shockwave and a dense circumstellar medium (CSM; see \citealt{filippenko1997} for a review). Based on the timescale of the observed shock interaction and the measured CSM wind velocities, significant mass loss must have taken place within about a decade of collapse, with mass loss rates  exceeding the line-driven wind maximum around $\sim 10^{-3} \mspy$ \citep[e.g.,][]{kiewe2012}. Some IIn also show evidence of continued  interaction signatures and light echoes from dust shells that imply elevated mass loss rates during or even before the carbon burning phase, $\gg 10 \yr$ prior to core collapse \citep{fox2011,kochanek2012b,fox2013}. 

Our wave-driven mass loss mechanism is broadly consistent with the early time observations of Type IIn SNe that imply high density CSM extending to $\sim 100{\rm s} \AU$.  Progenitors with the smallest helium core masses, $\lesssim 10 \msun$, are most likely to produce these types of CSM environments via waves driven by convective neon and oxygen burning (see Fig. \ref{fig:tonset} \& Table~\ref{tab:res}). However, our wave-driven mass loss mechanism cannot produce the more extended CSM that must result from winds launched more than a $\sim$ decade prior to core-collapse. These winds must be driven during core carbon burning (or even helium burning in some cases) when waves excited by core convection do not carry enough luminosity to significantly affect the stellar envelope.

The majority of massive stars are in close binaries, many of which will interact and undergo mass transfer during their lifetime \citep{sana2012}. As a result, it is likely that interaction between close binary companions also plays a role in the enhanced pre-explosion mass loss inferred in many Type IIn SNe (see \S \ref{sec:future}).

\subsubsection{Type IIb SNe}

\citet{chevalier2010} argued that Type IIb SNe appear to arise from both compact  ($R \sim R_\odot$) and much more extended progenitors ($R \gtrsim 100 R_\odot$).   Empirically, this division manifests itself as an approximate dichotomy in radio emission and early thermal SN emission, consistent with significant differences in the progenitor radius and the strength of the pre-SN wind.    SNe 1993j and  SNe 2011dh are examples of Type IIbs from extended progenitors \citep{aldering1994,bersten2012,vandyk2013}.    We suggest that this difference between compact and extended progenitors is caused by efficient wave energy deposition and ensuing mass loss in a subset of Type IIb progenitors.  The most plausible alternative interpretation  is that the hydrogen envelope masses are systematically smaller in the compact Type IIb progenitors \citep{chevalier2010}.   Tests of this hypothesis and/or evidence for (or limits on) pre-SN outbursts would provide tests of the importance of wave energy deposition in inflating the stellar radii of Type IIb SNe progenitors.

\subsubsection{Type Ibn: SN 2006jc} \label{sssec:2006jc}

Two years before its ultimate explosion in October 2006, an amateur astronomer recorded a luminous outburst at the same position as SN 2006jc \citep{foley2007,pastorello2007}. SN 2006jc is classified as Type Ibn, indicating that it lacks hydrogen in its spectrum (though some was detected at late times) and has relatively narrow lines.  The progenitor is plausibly a He-rich WR star, perhaps recently transitioned from an LBV phase (to explain the weak signatures of hydrogen at late times). 

While the precursor event was not well studied, the interaction between the SN ejecta and the pre-existing circumstellar material has enabled estimates of the mass-loss from this precursor. Through observations of x-rays (from the shock interaction), helium emission and p-cygni lines, and newly formed hot dust, several authors have estimated that the He-rich pre-SN ejecta has a velocity $\lesssim 2000 \kms$ and total ejecta mass $\gtrsim 10^{-2} \msun$ \citep{immler2008,smith2008,anupama2009}.   The inferred ejecta energetics, velocities, and masses are reasonably consistent with what we would expect from wave-driven mass loss in WR progenitors during neon and/or oxygen fusion (see Figs \ref{fig:oneEwave} \& \ref{fig:oneMej}).  However, none of the WR progenitors in our grid of stellar models produce wave-driven mass loss earlier than $\sim 6$ months prior to core collapse.   In particular,  we find that significant wave-driven mass loss 2 years prior to core collapse occurs only in stars with low He core masses, which are blue or red super-giants with extended hydrogen-rich envelopes in our calculations (Figs. \ref{fig:envmass} \& \ref{fig:tonset}).   We believe that the need for a low He core mass ($\sim 5 \msun$) to explain wave-driven mass loss 2 years prior to core collapse is robust but it is quite possible that the absence of WR progenitors at this He core mass in our grid of stellar models is simply a shortcoming of our mass loss prescriptions or our restriction to single star (vs. binary) evolution.

\subsubsection{Type IIn: SN 2009ip} \label{sssec:2009ip}

The SN 2009ip is distinguished by the multiple observed outbursts, which occur three years, one year and two months prior to core collapse (\citealt{mauerhan2013}; though its status as a ``true'' core-collapse event is still under debate, see \citealt{fraser2013}). Based on shock interaction with the ejecta, \citet{ofek2013a} estimate that $\sim 0.04 \msun$ of material was ejected over the last three years, at speeds of $\sim 10^{3} \kms$, reaching $\sim 6 \times 10^{15} \cm$ from the progenitor at the time of explosion. In pre-explosion imaging, \citet{mauerhan2013} find a blue, $10^{5.9} \lsun$ progenitor, from which they infer a $\sim 50 - 80 \msun$ ZAMS mass. 

As with SN 2006jc, the long pre-collapse outburst timescales for the first two outbursts of SN 2009ip would imply a low core mass in the wave-driven mass loss model, ${\rm M_{He}} \lesssim 5 \msun$. In our calculations these progenitors are an order of magnitude too faint at core collapse, relative to the pre-explosion images.   However, pre-explosion imaging  of stars that undergo dramatic late-time mass loss may already catch the star in a state that is not well-described by any existing stellar evolutionary models (this worry is less acute for more typical Type IIp progenitors).

\subsubsection{Type IIn: SN 2010mc/PTF 10tel} \label{sssec:2010mc}

The pre-SN outburst from SN 2010mc, which occurred 40 days prior to core collapse, matches our models closest of the three observed examples (as argued by \citealt{ofek2013}). The precursor for this SN radiated $\sim 6 \times 10^{47} \erg$ and expelled $\gtrsim 10^{-2} \msun$ at speeds of $\sim 2000 \kms$. The 40 \Msun{} sub-solar metallicity progenitors we have simulated generate comparable wave-driven events during core oxygen-burning.

\subsection{Directions for future work} \label{sec:future}

The conditions we have investigated in this paper are necessary for wave-driven mass loss during late stages of stellar evolution, but it not yet clear if they are sufficient. Further calculations, including hydrodynamic simulations like those of \citet{meakin2006}, are necessary to investigate the excitation, propagation, and damping of waves in SN progenitors. Both multi-dimensional hydrodynamic simulations \citep[like those by, e.g.,][]{rogers2006a,browning2004} and observations (e.g., \citealt{shiode2013}) would provide valuable constraints on the spectrum of wave excitation and thus how much of the full wave energy reservoir can reach the stellar envelope.    The principal uncertainty is that only low spherical harmonic degree modes ($\ell \lesssim$ a few) can efficiently tunnel from the stellar core to the envelope, so the energetics of wave-driven mass loss depends critically on the fraction of the internal gravity wave power in low degree modes.   This in turn likely depends on the convective mach number, the stellar rotation rate (specifically the Rossby number of the convection) and the structure of the convective-radiative transition region.   It is  possible that {\em no} stellar progenitors excite sufficiently low $\ell$ modes in late stages of stellar evolution to power significant mass loss.

The restriction to low $\ell$ modes is also the reason that we have focused on wave excitation by core fusion rather than shell fusion (though Figs \ref{fig:oneEwave} \& \ref{fig:siEwave} quantify the energetics of shell fusion in our progenitors).    Wave excitation by shell fusion is more likely to excite high $\ell$ modes  because the size of the convective eddies are limited by the thickness of the shell.   On the other hand, waves excited by shell fusion have less of a `barrier' to tunnel through to reach the stellar envelope.   Our estimates suggest that core fusion is nonetheless the most promising site for waves capable of powering significant mass loss, but a more definitive conclusion on this question will require fully understanding the power-spectrum of waves excited by stellar convection.

Multi-dimensional hydrodynamical simulations are also needed to better understand the behavior of super-Eddington stellar envelopes. Some authors, including \citet{soker2013}, have suggested that envelope inflation is a more likely outcome than mass ejection or the formation of a super-Eddington wind, which we favor \citep[][have also argued for the latter]{shaviv2001,owocki2004}.   It is  unclear how critical this distinction is:   SN ejecta interacting with a significantly inflated (but bound) stellar envelope will produce emission via ``circumstellar" interaction similar to that produced by interaction with an unbound outflow.      

Since most massive stars are in binary systems, a clear area for future research is the interplay between wave-driven mass loss (and/or radius inflation) and Roche-lobe overflow in close binary systems (see \citealt{soker2013}).  Mass transfer in a binary may determine the radius of the progenitor at core collapse and influence how much mass is ejected from the binary system during episodes of efficient wave energy deposition in the stellar envelope.

Some of our progenitors are susceptible to wave-driven mass loss at multiple stages during their evolution to collapse. However, at each phase, we have investigated our 1-D stellar evolution model without any enhanced mass loss during prior phases. Future efforts to quantify the potential effect of wave-driven mass loss on the subsequent evolution of a star are necessary to understand the full evolution of progenitors that experience pre-SN outbursts. It is unclear how wave-driven mass loss events would affect the future evolution of the stellar core and envelope, and thus any potential further wave-driven mass loss.

Throughout our investigation, we have ignored the effect of rotation on the excitation, propagation and damping of waves.  This is likely to be a rather poor approximation in many cases, especially for our rapidly rotating progenitors. The excitation of modes depends on the statistical properties of convection, which are different in rapidly rotating stars. Rotation also modifies the efficiency of chemical mixing, affects the shape of wave propagation cavities, and can introduce critical damping layers \citep[e.g.,][]{rogers2012} that might inhibit waves from reaching the stellar surface.   We will investigate these  effects in future work.

\section*{Acknowledgments}

We  thank Bill Paxton for invaluable support with MESA. We also thank David Arnett, Lars Bildsten, Matteo Cantiello, Dan Kasen, Tony Piro, and Nathan Smith for fruitful discussions. JS would like to thank the organizers and attendees of the Workshop on Outstanding Problems in Massive Star Research - The Final Stages in Minneapolis, MN in 2012 for useful discussions.  This work was partially supported by a Simons Investigator award from the Simons Foundation to EQ, the David and Lucile Packard Foundation, and the Thomas and Alison Schneider Chair in Physics at UC Berkeley.  Further support was provided by NASA Headquarters under the NASA Earth and Space Science Fellowship Program - Grant 10-Astro10F-0030.

\appendix

\section*{Appendix A: Massive star models} \label{app:inlist}

We use version 4789 of the \MESAstar{} stellar evolution code \citep{mesa2011,mesaII} to construct evolutionary sequences of massive stars from the zero-age main sequence (ZAMS) to core collapse. We employed four separate inlists to evolve each progenitor from start to finish.\footnote{These will be made available, in full, on http://mesastar.org.} The first generates ZAMS models. For masses below 30 \Msun{}, we use the following non-default parameters
{\small \begin{lstlisting}
         create_pre_main_sequence_model = .true.
         mesh_delta_coeff = 0.5
         Lnuc_div_L_upper_limit = 0.9
         overshoot_f_above_burn_h = 0.335
         overshoot_f0_above_burn_h = 0.
         overshoot_step_fraction = 1.
\end{lstlisting}} \noindent with 
{\small \begin{lstlisting}
         relax_Z = .true.
         new_Z = <value>
\end{lstlisting}}
\noindent for non-solar metallicities and 
{\small \begin{lstlisting}
         change_rotation_flag = .true.
         new_rotation_flag = .true.
         set_omega_div_omega_crit = .true.
         new_omega_div_omega_crit = <value>
\end{lstlisting}}
\noindent for rotating models.   Above 30 \Msun{}, we read in an analogous (in terms of rotation and metallicity) 30 \Msun{} model stopped at \lstinline{Lnuc_div_L_upper_limit = 0.1} and use 
{\small \begin{lstlisting}
         relax_mass_scale = .true.
         new_mass = <value>
\end{lstlisting}}
\noindent instead of \lstinline{create_pre_main_sequence_model = .true.}.    The next inlist evolves the ZAMS models through the main sequence using
{\small \begin{lstlisting}
         change_v_flag = .true.
         new_v_flag = .true.
         set_rate_c12ag = 'Kunz'
         set_rate_n14pg = 'Imbriani'
         set_rate_3a = 'Fynbo'
         kappa_file_prefix = 'gs98' 
         mesh_delta_coeff = 0.5
         use_Type2_opacities = .true.
         mixing_length_alpha = 1.5
         use_Henyey_MLT = .true.
         use_Ledoux_criterion = .true.
         alpha_semiconvection = 0.1
         thermo_haline_coeff = 2.0
         T_mix_limit = 0
         max_iter_for_resid_tol1 = 3
         tol_residual_norm1 = 1d-5
         max_tries = 50
         max_tries_for_retry = 50
         max_tries_after_backup = 50
         max_tries_after_backup2 = 50
         delta_lgL_He_limit = -1
         delta_lgP_limit = -1
         delta_lgTeff_limit = 0.5 
         delta_lgL_limit = 0.5 
         delta_lgRho_cntr_limit = 0.02 
         dX_nuc_drop_limit = 5d-3
\end{lstlisting}}
\noindent with overshoot above the H-burning core turned on as during the pre-MS evolution above. The envelope is allowed to more efficiently mix using the ``enhanced MLT'' scheme with \lstinline{okay_to_reduce_gradT_excess = .true.} and velocities are limited to the stellar core using 
{\small \begin{lstlisting}
         velocity_logT_lower_bound = 9 
         velocity_Z_lower_bound =  10
\end{lstlisting}}
\noindent Resolution is increased by 0.4 in all transition regions. The mass loss schemes are either 
{\small \begin{lstlisting}
         RGB_wind_scheme = 'Dutch'
         AGB_wind_scheme = 'Dutch'
         Dutch_wind_lowT_scheme = 'de Jager'
         RGB_to_AGB_wind_switch = 1d-4
\end{lstlisting}}
\noindent or 
{\small \begin{lstlisting}
         RGB_wind_scheme = 'Dutch'
         AGB_wind_scheme = 'Dutch'
         Dutch_wind_lowT_scheme = 'Nieuwenhuijzen'
         RGB_to_AGB_wind_switch = 1d-4
\end{lstlisting}}
\noindent with \lstinline{Dutch_wind_eta} set as described in \S\ref{sec:mods}. For rotating models, the diffusion coefficients for rotational mixing are set as the following
{\small \begin{lstlisting}
         D_SH_factor =  0.0
         D_SSI_factor = 1.16
         D_ES_factor =  1.16
         D_GSF_factor = 1.16
\end{lstlisting}}
\noindent The models are stopped when they reach the end of the MS, controlled by the parameters
{\small \begin{lstlisting}
         xa_central_lower_limit_species(1) = 'h1'
         xa_central_lower_limit(1) = 1e-8
\end{lstlisting}}

\noindent Beyond the MS, we turn off the overshoot by setting
{\small \begin{lstlisting}
         overshoot_f_above_burn_h = 0.0
         overshoot_f0_above_burn_h = 0.
         overshoot_step_fraction = 0.
\end{lstlisting}}
\noindent and evolve the models to central carbon exhaustion at 
{\small \begin{lstlisting}
         xa_central_lower_limit_species(1) = 'c12'
         xa_central_lower_limit(1) = 1e-6
\end{lstlisting}}
\noindent Finally, we run the models from neon and oxygen burning to core collapse at lower resolution by setting \lstinline{mesh_delta_coeff = 1.0}.

This procedure works uninterrupted for the majority of our model grid. However, in some cases it was necessary to change a few of \MESAstar's other parameters to aid convergence. For massive stars in which the envelope convection zone reaches deep down towards the hydrogen burning shell during post-ms evolution, we sometimes needed to increase \lstinline{dH_div_H_limit} to allow relatively large changes in fractional hydrogen abundance due to the changing mesh at the convective boundary. In a few cases, we needed to increase solver tolerances (\lstinline{tol_correction_norm} and \lstinline{tol_max_correction}) to ten or 30 times the default during the final evolutionary phases to reach core collapse.

\vspace{1cm}

\clearpage

\section*{Appendix B:  Model Parameters and Results}   \label{app:tables}

\setlength{\tabcolsep}{0.045in}


\begin{center}
\begin{longtable*}{cccccrrrrrrrrrrr}
   \caption{Stellar model properties} \label{tab:mods}\\
    
   \\[-4ex]
   \hline \hline \\[-2ex]
       ${\rm M_{ZAMS}}$ & ${\rm Z_{init}}$ & $\Omega/\Omega_{\rm crit}^{\ast}$ & Mass Loss$^{\ast \ast}$ & 
       Class$^{\dagger}$ & ${\rm M_{He}}^{\ddagger}$ & ${\rm M_{C/O}}$ & ${\rm M_{iron}}$ & 
       ${\rm M_{env, \, H}}$ & ${\rm M_{env, \, He}}$ & \multicolumn{6}{c}{Convective (radiative) core burning time$^{\ast \ast \ast}$}\\
       \cline{11-16}\\[-1.8ex]
       $\left[{\rm M}_{\odot}\right]$ & $\left[{\rm Z}_\odot\right]$ & & & & $\left[{\rm M}_{\odot}\right]$ & 
       $\left[{\rm M}_{\odot}\right]$ & $\left[{\rm M}_{\odot}\right]$ & 
       $\left[{\rm M}_{\odot}\right]$ & $\left[{\rm M}_{\odot}\right]$ & H [Myr] & He [Myr] & C [yr] & Ne [yr] & O [yr] & Si [day]\\[0.5ex]
   \hline \\[-1.8ex]
   \endfirsthead
    
\multicolumn{16}{c}{{\tablename} \thetable{} --- Continued} \\
   \hline \hline \\[-2ex]
       ${\rm M_{ZAMS}}$ & ${\rm Z_{init}}$ & $\Omega/\Omega_{\rm crit}^{\ast}$ & Mass Loss$^{\ast \ast}$ & 
       Class$^{\dagger}$ & ${\rm M_{He}}^{\ddagger}$ & ${\rm M_{C/O}}$ & ${\rm M_{iron}}$ & 
       ${\rm M_{env, \, H}}$ & ${\rm M_{env, \, He}}$ & \multicolumn{6}{c}{Convective (radiative) core burning time$^{\ast \ast \ast}$}\\
       \cline{11-16}\\[-1.8ex]
       $\left[{\rm M}_{\odot}\right]$ & $\left[{\rm Z}_\odot\right]$ & & & & $\left[{\rm M}_{\odot}\right]$ & 
       $\left[{\rm M}_{\odot}\right]$ & $\left[{\rm M}_{\odot}\right]$ & 
       $\left[{\rm M}_{\odot}\right]$ & $\left[{\rm M}_{\odot}\right]$ & H [Myr] & He [Myr] & C [yr] & Ne [yr] & O [yr] & Si [day]\\[0.5ex]
   \hline \\[-1.8ex]
   \endhead
    
       \\[-2ex]
        \hline \hline\\
        \multicolumn{16}{l}{Continued on Next Page\ldots}
   \endfoot
    
       \\[-2ex]
       \hline \hline
	   \\
       \multicolumn{16}{l}{$^{\ast}\Omega_{\rm crit}^2 \equiv (1 - L/\Ledd) G M_\star / R_\star^3$.}\\
       \multicolumn{16}{l}{$^{\ast \ast}$Here, ``vdj'' is shorthand for the \citet{vink2001} and \citet{de-jager1988} mass loss prescription combination and the dimensionless efficiency factor is}\\
       \multicolumn{16}{l}{shown as a prefactor; ``v+n'' denotes the \citet{vink2001} and \citet{nieu1990} mass loss prescription combination.}\\
       \multicolumn{16}{l}{$^{\dagger}$We define ``RSG'' as a progenitor that retains its hydrogen envelope and has $T_{\rm eff} < 10^{4} \K$ at core collapse, ``BSG'' as progenitor with a hydrogen envelope}\\
       \multicolumn{16}{l}{and $T_{\rm eff} \geq 10^{4} \K$, and a ``WR'' as a progenitor without hydrogen.}\\
       \multicolumn{16}{l}{$^{\ddagger}$Helium core mass at core infall; same as the ``He core mass'' plotted on the abcissa of figs.~\ref{fig:envmass}, \ref{fig:oneMej}, \ref{fig:siMej}, and \ref{fig:tonset}}\\
       \multicolumn{16}{l}{$^{\ast \ast \ast}$Burning lifetimes are determined based on the time it takes for burning to reduce the central abundance of the main fuel below $10^{-5}$. Except for Ne, }\\
       \multicolumn{16}{l}{which does not fully deplete before core oxygen burning, for which we estimate the lifetime based on the start of Ne burning to the start of O burning. Note that }\\
       \multicolumn{16}{l}{the lifetime of the central convection zone, when one exists, may be shorter than the lifetime of the burning phase given here (integrated energies relevant for }\\
		\multicolumn{16}{l}{excited waves span only the convective lifetime).}
   \endlastfoot
      $12$ & $0$ & $0.00$ & 0.8 (v+dj) & RSG & $3.17$ & $1.98$ & $1.51$ & $5.89$ & $3.98$ & $15.0$ & $0.85$ & $6600.0$ & $6.300$ & $5.800$ & $4.90$\\
      $12$ & $10^{-2}$ & $0.00$ & 0.8 (v+dj) & RSG & $4.39$ & $2.14$ & $1.49$ & $5.31$ & $4.14$ & $19.0$ & $1.10$ & $3800.0$ & $4.300$ & $4.000$ & $3.60$\\
      $12$ & $10^{-1}$ & $0.00$ & 0.8 (v+dj) & RSG & $4.35$ & $2.11$ & $1.45$ & $4.49$ & $3.89$ & $19.0$ & $1.10$ & $4900.0$ & $3.600$ & $4.300$ & $4.40$\\
      $12$ & $1.0$ & $0.00$ & 0.8 (v+dj) & RSG & $4.21$ & $1.97$ & $1.44$ & $3.99$ & $3.85$ & $18.0$ & $1.10$ & $5600.0$ & $4.600$ & $5.000$ & $5.50$\\
      $12$ & $1.0$ & $0.20$ & 0.6 (v+dj) & RSG & $4.94$ & $2.34$ & $1.46$ & $3.75$ & $4.00$ & $16.0$ & $0.99$ & $3100.0$ & $3.700$ & $2.800$ & $3.60$\\
      $12$ & $1.0$ & $0.50$ & 0.6 (v+dj) & RSG & $2.79$ & $2.36$ & $1.53$ & $3.54$ & $4.00$ & $17.0$ & $0.97$ & $2900.0$ & $6.100$ & $3.400$ & $3.70$\\
      $12$ & $1.0$ & $0.80$ & 0.6 (v+dj) & RSG & $6.57$ & $3.20$ & $1.49$ & $1.86$ & $4.78$ & $21.0$ & $0.75$ & $1300.0$ & $0.780$ & $3.100$ & $2.80$\\[0.5ex]
      \hline\\[-2ex]
      $15$ & $0$ & $0.00$ & 0.8 (v+dj) & BSG & $3.62$ & $2.82$ & $1.50$ & $6.98$ & $4.87$ & $12.0$ & $0.66$ & $2300.0$ & $1.000$ & $3.900$ & $2.80$\\
      $15$ & $10^{-2}$ & $0.00$ & 0.8 (v+dj) & RSG & $5.84$ & $2.83$ & $1.47$ & $6.26$ & $5.09$ & $14.0$ & $0.80$ & $1800.0$ & $4.300$ & $2.100$ & $2.70$\\
      $15$ & $10^{-1}$ & $0.00$ & 0.8 (v+dj) & RSG & $5.82$ & $2.80$ & $1.47$ & $4.50$ & $4.51$ & $14.0$ & $0.81$ & $1900.0$ & $3.300$ & $2.200$ & $2.30$\\
      $15$ & $1.0$ & $0.00$ & 0.8 (v+dj) & RSG & $5.72$ & $2.67$ & $1.48$ & $3.67$ & $4.27$ & $13.0$ & $0.82$ & $2100.0$ & $4.100$ & $2.200$ & $3.00$\\
      $15$ & $1.0$ & $0.50$ & 0.6 (v+dj) & RSG & $6.69$ & $3.26$ & $1.58$ & $3.45$ & $4.53$ & $12.0$ & $0.73$ & $1200.0$ & $3.500$ & $2.300$ & $2.10$\\
      $15$ & $1.0$ & $0.80$ & 0.6 (v+dj) & RSG & $9.48$ & $4.69$ & $1.65$ & $0.80$ & $4.54$ & $16.0$ & $0.59$ & $(230.0)$ & $0.660$ & $0.700$ & $0.89$\\[0.5ex]
      \hline\\[-2ex]
      $20$ & $0$ & $0.00$ & 0.8 (v+dj) & BSG & $5.24$ & $4.45$ & $1.46$ & $8.44$ & $6.55$ & $8.7$ & $0.50$ & $650.0$ & $0.520$ & $1.600$ & $3.50$\\
      $20$ & $10^{-2}$ & $0.00$ & 0.8 (v+dj) & RSG & $7.78$ & $4.40$ & $1.59$ & $6.99$ & $6.45$ & $9.9$ & $0.55$ & $560.0$ & $0.290$ & $0.420$ & $1.10$\\
      $20$ & $10^{-1}$ & $0.00$ & 0.8 (v+dj) & RSG & $8.49$ & $4.26$ & $1.53$ & $3.21$ & $5.06$ & $9.9$ & $0.58$ & $540.0$ & $0.910$ & $1.600$ & $2.30$\\
      $20$ & $1.0$ & $0.00$ & 0.8 (v+dj) & RSG & $8.45$ & $4.27$ & $1.48$ & $2.03$ & $4.53$ & $9.0$ & $0.57$ & $670.0$ & $1.900$ & $1.900$ & $1.70$\\
      $20$ & $1.0$ & $0.20$ & 0.6 (v+dj) & RSG & $7.70$ & $5.12$ & $1.64$ & $2.71$ & $5.04$ & $8.5$ & $0.55$ & $(330.0)$ & $0.640$ & $1.300$ & $1.60$\\
      $20$ & $1.0$ & $0.50$ & 0.6 (v+dj) & RSG & $9.70$ & $5.30$ & $1.48$ & $2.17$ & $4.93$ & $8.7$ & $0.54$ & $(340.0)$ & $0.260$ & $0.770$ & $2.20$\\
      $20$ & $1.0$ & $0.80$ & 0.6 (v+dj) & WR & $10.96$ & $7.47$ & $1.85$ & $-$ & $2.21$ & $11.0$ & $0.48$ & $(150.0)$ & $1.200$ & $0.380$ & $0.65$\\[0.5ex]
      \hline\\[-2ex]
      $25$ & $0$ & $0.00$ & 0.8 (v+dj) & BSG & $9.41$ & $6.73$ & $1.57$ & $9.84$ & $7.86$ & $7.0$ & $0.43$ & $(280.0)$ & $0.140$ & $0.360$ & $1.30$\\
      $25$ & $10^{-2}$ & $0.00$ & 0.8 (v+dj) & RSG & $11.05$ & $6.65$ & $1.83$ & $2.67$ & $6.18$ & $7.8$ & $0.46$ & $(180.0)$ & $0.910$ & $0.300$ & $0.55$\\
      $25$ & $10^{-1}$ & $0.00$ & 0.8 (v+dj) & RSG & $11.34$ & $6.43$ & $1.64$ & $1.38$ & $5.49$ & $7.8$ & $0.46$ & $(240.0)$ & $0.110$ & $0.230$ & $0.95$\\
      $25$ & $1.0$ & $0.00$ & 0.8 (v+dj) & RSG & $11.16$ & $6.48$ & $1.82$ & $0.85$ & $4.71$ & $7.1$ & $0.47$ & $(210.0)$ & $0.160$ & $0.150$ & $0.64$\\
      $25$ & $1.0$ & $0.20$ & 0.6 (v+dj) & RSG & $12.39$ & $7.39$ & $1.63$ & $0.97$ & $5.04$ & $6.8$ & $0.47$ & $82.0$ & $0.620$ & $0.670$ & $0.82$\\
      $25$ & $1.0$ & $0.50$ & 0.6 (v+dj) & RSG & $13.07$ & $8.11$ & $1.64$ & $0.55$ & $4.69$ & $7.2$ & $0.45$ & $62.0$ & $0.450$ & $0.590$ & $0.89$\\
      $25$ & $1.0$ & $0.80$ & 0.6 (v+dj) & WR & $11.36$ & $8.67$ & $1.79$ & $-$ & $1.29$ & $8.9$ & $0.47$ & $(110.0)$ & $1.300$ & $0.390$ & $0.77$\\[0.5ex]
      \hline\\[-2ex]
      $30$ & $0$ & $0.00$ & 0.8 (v+dj) & BSG & $11.90$ & $8.50$ & $1.68$ & $11.10$ & $9.83$ & $6.0$ & $0.37$ & $(110.0)$ & $0.360$ & $0.530$ & $0.74$\\
      $30$ & $10^{-2}$ & $0.00$ & 0.8 (v+dj) & RSG & $13.98$ & $9.11$ & $1.53$ & $1.18$ & $6.24$ & $6.6$ & $0.41$ & $(100.0)$ & $0.490$ & $0.660$ & $1.90$\\
      $30$ & $10^{-1}$ & $0.00$ & 0.8 (v+dj) & RSG & $14.28$ & $8.81$ & $1.55$ & $0.85$ & $6.31$ & $6.6$ & $0.40$ & $(100.0)$ & $0.320$ & $0.490$ & $1.00$\\
      $30$ & $1.0$ & $0.00$ & 0.8 (v+dj) & WR & $13.86$ & $8.07$ & $1.64$ & $-$ & $4.24$ & $6.0$ & $0.41$ & $(82.0)$ & $0.300$ & $0.490$ & $1.10$\\
      $30$ & $1.0$ & $0.20$ & 0.6 (v+dj) & BSG & $15.46$ & $9.65$ & $1.78$ & $0.45$ & $5.29$ & $5.8$ & $0.42$ & $37.0$ & $0.120$ & $0.190$ & $0.91$\\
      $30$ & $1.0$ & $0.50$ & 0.6 (v+dj) & BSG & $16.33$ & $10.49$ & $1.86$ & $0.35$ & $5.16$ & $6.1$ & $0.41$ & $(58.0)$ & $(0.100)$ & $0.160$ & $0.76$\\
      $30$ & $1.0$ & $0.80$ & 0.6 (v+dj) & WR & $12.65$ & $9.67$ & $1.52$ & $-$ & $0.96$ & $7.4$ & $0.44$ & $60.0$ & $0.750$ & $0.870$ & $2.50$\\[0.5ex]
      \hline\\[-2ex]
      $40$ & $0$ & $0.00$ & 0.8 (v+dj) & RSG & $17.48$ & $14.49$ & $1.85$ & $12.57$ & $11.09$ & $4.8$ & $0.36$ & $(26.0)$ & $(0.079)$ & $0.095$ & $0.69$\\
      $40$ & $10^{-2}$ & $0.00$ & 0.8 (v+dj) & RSG & $20.02$ & $14.11$ & $1.87$ & $1.42$ & $8.44$ & $5.2$ & $0.34$ & $(24.0)$ & $(0.078)$ & $0.093$ & $0.59$\\
      $40$ & $10^{-1}$ & $0.00$ & 0.8 (v+dj) & BSG & $20.37$ & $13.98$ & $1.97$ & $0.85$ & $7.84$ & $5.2$ & $0.35$ & $(34.0)$ & $(0.072)$ & $0.093$ & $0.62$\\
      $40$ & $1.0$ & $0.00$ & v+n & WR & $10.04$ & $7.63$ & $1.73$ & $-$ & $0.31$ & $4.8$ & $0.43$ & $(170.0)$ & $0.240$ & $0.180$ & $0.73$\\
      $40$ & $1.0$ & $0.00$ & 0.8 (v+dj) & WR & $13.10$ & $10.24$ & $1.51$ & $-$ & $0.32$ & $4.8$ & $0.39$ & $(67.0)$ & $0.410$ & $0.490$ & $1.50$\\
      $40$ & $1.0$ & $0.20$ & 0.6 (v+dj) & WR & $20.41$ & $14.77$ & $1.67$ & $-$ & $3.81$ & $4.7$ & $0.36$ & $(15.0)$ & $(0.061)$ & $0.067$ & $1.50$\\
      $40$ & $1.0$ & $0.50$ & 0.6 (v+dj) & WR & $20.01$ & $15.61$ & $2.10$ & $-$ & $2.55$ & $4.9$ & $0.36$ & $9.8$ & $(0.058)$ & $0.068$ & $1.10$\\
      $40$ & $1.0$ & $0.80$ & 0.6 (v+dj) & WR & $13.77$ & $11.22$ & $1.58$ & $-$ & $0.21$ & $5.7$ & $0.41$ & $(41.0)$ & $0.380$ & $0.360$ & $1.50$\\[0.5ex]
      \hline\\[-2ex]
      $50$ & $0$ & $0.00$ & 0.8 (v+dj) & RSG & $23.13$ & $19.60$ & $1.75$ & $9.97$ & $12.21$ & $4.2$ & $0.33$ & $(11.0)$ & $(0.040)$ & $0.044$ & $1.00$\\
      $50$ & $10^{-2}$ & $0.00$ & 0.8 (v+dj) & BSG & $26.28$ & $20.05$ & $1.74$ & $1.91$ & $10.01$ & $4.5$ & $0.32$ & $(13.0)$ & $(0.041)$ & $0.046$ & $1.10$\\
      $50$ & $10^{-1}$ & $0.00$ & 0.8 (v+dj) & BSG & $26.61$ & $19.40$ & $1.75$ & $0.85$ & $8.40$ & $4.5$ & $0.33$ & $(17.0)$ & $(0.042)$ & $0.046$ & $0.89$\\
      $50$ & $1.0$ & $0.00$ & 0.8 (v+dj) & WR & $12.35$ & $9.44$ & $1.67$ & $-$ & $0.34$ & $4.2$ & $0.40$ & $(78.0)$ & $0.570$ & $0.650$ & $2.00$\\
      $50$ & $1.0$ & $0.00$ & v+n & WR & $10.42$ & $7.99$ & $1.83$ & $-$ & $0.30$ & $4.2$ & $0.41$ & $(150.0)$ & $1.100$ & $0.330$ & $0.63$\\
      $50$ & $1.0$ & $0.20$ & 0.6 (v+dj) & WR & $24.19$ & $20.15$ & $1.76$ & $-$ & $2.17$ & $4.1$ & $0.34$ & $(10.0)$ & $(0.031)$ & $0.038$ & $0.86$\\
      $50$ & $1.0$ & $0.50$ & 0.6 (v+dj) & WR & $22.33$ & $16.34$ & $1.74$ & $-$ & $0.44$ & $4.2$ & $0.34$ & $(9.8)$ & $(0.037)$ & $0.044$ & $1.00$\\
      $50$ & $1.0$ & $0.80$ & 0.6 (v+dj) & WR & $14.36$ & $11.65$ & $1.59$ & $-$ & $0.25$ & $4.9$ & $0.40$ & $(28.0)$ & $(0.340)$ & $0.220$ & $1.30$\\[0.5ex]
      \hline\\
      $60$ & $0$ & $0.00$ & 0.8 (v+dj) & BSG & $29.15$ & $25.39$ & $2.27$ & $3.16$ & $11.45$ & $3.8$ & $0.31$ & $(9.6)$ & $(0.024)$ & $0.028$ & $0.28$\\
      $60$ & $10^{-2}$ & $0.00$ & 0.8 (v+dj) & BSG & $32.60$ & $25.34$ & $1.99$ & $1.98$ & $11.42$ & $4.0$ & $0.31$ & $(7.0)$ & $(0.026)$ & $0.026$ & $0.52$\\
      $60$ & $10^{-1}$ & $0.00$ & 0.8 (v+dj) & BSG & $32.75$ & $24.84$ & $2.20$ & $0.13$ & $7.18$ & $4.1$ & $0.31$ & $(7.8)$ & $(0.025)$ & $0.024$ & $0.31$\\
      $60$ & $1.0$ & $0.00$ & v+n & WR & $11.70$ & $9.08$ & $1.57$ & $-$ & $0.32$ & $3.8$ & $0.39$ & $(81.0)$ & $0.770$ & $0.860$ & $2.40$\\
      $60$ & $1.0$ & $0.00$ & 0.8 (v+dj) & WR & $15.14$ & $12.09$ & $1.71$ & $-$ & $0.33$ & $3.8$ & $0.37$ & $(38.0)$ & $(0.200)$ & $0.170$ & $0.81$\\
      $60$ & $1.0$ & $0.20$ & 0.6 (v+dj) & WR & $26.93$ & $21.57$ & $2.12$ & $-$ & $0.52$ & $3.7$ & $0.32$ & $(4.9)$ & $(0.020)$ & $0.029$ & $0.39$\\
      $60$ & $1.0$ & $0.50$ & 0.6 (v+dj) & WR & $24.56$ & $20.61$ & $1.81$ & $-$ & $0.44$ & $3.8$ & $0.33$ & $(7.9)$ & $(0.033)$ & $0.040$ & $0.65$\\
      $60$ & $1.0$ & $0.80$ & 0.6 (v+dj) & WR & $15.67$ & $12.74$ & $1.76$ & $-$ & $0.33$ & $4.3$ & $0.39$ & $(24.0)$ & $(0.220)$ & $0.150$ & $0.78$\\[0.5ex]
      \hline\\[-2ex]
      $70$ & $0$ & $0.00$ & 0.8 (v+dj) & BSG & $34.99$ & $30.95$ & $2.63$ & $3.60$ & $13.42$ & $3.5$ & $0.30$ & $(5.2)$ & $(0.017)$ & $0.019$ & $0.24$\\
      $70$ & $10^{-2}$ & $0.00$ & 0.8 (v+dj) & BSG & $38.95$ & $31.00$ & $2.54$ & $2.35$ & $13.18$ & $3.7$ & $0.29$ & $(5.8)$ & $(0.019)$ & $0.017$ & $0.25$\\
      $70$ & $10^{-1}$ & $0.00$ & 0.8 (v+dj) & BSG & $38.97$ & $30.89$ & $2.58$ & $0.05$ & $7.77$ & $3.7$ & $0.29$ & $(4.9)$ & $(0.018)$ & $0.018$ & $0.24$\\
      $70$ & $1.0$ & $0.00$ & v+n & WR & $13.54$ & $10.65$ & $1.63$ & $-$ & $0.31$ & $3.5$ & $0.37$ & $(57.0)$ & $0.300$ & $0.250$ & $1.20$\\
      $70$ & $1.0$ & $0.00$ & 0.8 (v+dj) & WR & $17.45$ & $14.20$ & $1.87$ & $-$ & $0.35$ & $3.5$ & $0.35$ & $(22.0)$ & $(0.100)$ & $0.097$ & $0.62$\\
      $70$ & $1.0$ & $0.20$ & 0.6 (v+dj) & WR & $24.21$ & $20.24$ & $1.70$ & $-$ & $0.41$ & $3.4$ & $0.33$ & $(4.9)$ & $(0.035)$ & $0.043$ & $0.94$\\
      $70$ & $1.0$ & $0.50$ & 0.6 (v+dj) & WR & $25.68$ & $18.53$ & $1.97$ & $-$ & $0.41$ & $3.5$ & $0.33$ & $(4.3)$ & $(0.028)$ & $0.039$ & $0.38$\\
      $70$ & $1.0$ & $0.80$ & 0.6 (v+dj) & WR & $16.43$ & $13.36$ & $1.78$ & $-$ & $0.38$ & $3.9$ & $0.38$ & $(21.0)$ & $(0.180)$ & $0.130$ & $0.88$\\[0.5ex]
      \hline\\[-2ex]
      $80$ & $1.0$ & $0.00$ & 0.8 (v+dj) & WR & $18.89$ & $15.23$ & $2.01$ & $-$ & $0.31$ & $3.3$ & $0.34$ & $(19.0)$ & $(0.074)$ & $0.075$ & $0.93$\\
      $80$ & $1.0$ & $0.00$ & v+n & WR & $8.45$ & $6.22$ & $1.51$ & $-$ & $0.32$ & $3.3$ & $0.44$ & $290.0$ & $1.500$ & $1.800$ & $2.20$\\
      $80$ & $1.0$ & $0.20$ & 0.6 (v+dj) & WR & $19.83$ & $16.31$ & $2.13$ & $-$ & $0.40$ & $3.3$ & $0.36$ & $(9.1)$ & $(0.080)$ & $0.075$ & $0.78$\\
      $80$ & $1.0$ & $0.50$ & 0.6 (v+dj) & WR & $18.61$ & $15.11$ & $2.08$ & $-$ & $0.40$ & $3.3$ & $0.36$ & $(14.0)$ & $(0.098)$ & $0.086$ & $0.64$\\
      $80$ & $1.0$ & $0.80$ & 0.6 (v+dj) & WR & $17.54$ & $14.26$ & $2.04$ & $-$ & $0.39$ & $3.6$ & $0.37$ & $(16.0)$ & $(0.130)$ & $0.110$ & $0.72$\\[0.5ex]
      \hline\\[-2ex]
      $100$ & $1.0$ & $0.00$ & 0.8 (v+dj) & WR & $9.81$ & $7.45$ & $1.66$ & $-$ & $0.34$ & $3.0$ & $0.43$ & $(170.0)$ & $0.130$ & $0.200$ & $0.96$\\
      $100$ & $1.0$ & $0.00$ & v+n & WR & $6.34$ & $4.49$ & $1.57$ & $-$ & $0.34$ & $3.1$ & $0.52$ & $850.0$ & $0.740$ & $0.690$ & $1.40$\\
      $100$ & $1.0$ & $0.20$ & 0.6 (v+dj) & WR & $19.01$ & $13.14$ & $2.08$ & $-$ & $0.40$ & $3.0$ & $0.36$ & $(11.0)$ & $(0.100)$ & $0.089$ & $0.65$\\
      $100$ & $1.0$ & $0.50$ & 0.6 (v+dj) & WR & $18.06$ & $14.73$ & $2.04$ & $-$ & $0.40$ & $3.3$ & $0.37$ & $(10.0)$ & $(0.130)$ & $0.100$ & $0.60$\\
      $100$ & $1.0$ & $0.80$ & 0.6 (v+dj) & WR & $17.47$ & $14.25$ & $1.95$ & $-$ & $0.40$ & $3.5$ & $0.37$ & $(22.0)$ & $(0.140)$ & $0.110$ & $0.79$\\
\end{longtable*}

\end{center}



\setlength{\tabcolsep}{6pt}

\clearpage
\LongTables 

\begin{landscape}
	
\setlength{\tabcolsep}{0.025in}


\begin{center}
\begin{longtable}{ccccccrrrrr|rrrrr|rrrrr}
   \caption{Wave-driven mass loss results} \label{tab:res}\\
    
   \\[-4ex]
   \hline \hline \\[-2ex]
      & & & & & & \multicolumn{5}{c}{\underline{core Ne burning}} & 
      \multicolumn{5}{c}{\underline{core O burning}} & \multicolumn{5}{c}{\underline{core Si burning}}\\
      \\[-1.8ex]
      ${\rm M_{ZAMS}}$ & ${\rm Z_{init}}$ & $\Omega/\Omega_{\rm crit}$ & Mass Loss & Class & ${\rm M_{He}}$ &
      $\tcc^\ast$ & ${\rm E_{w, \, 46}}^{\dagger}$ & ${\rm M_{ej}}^{\ddagger}$ & $v_{\rm esc}^{\ddagger}$ & ${\rm R_{ej}}^{\ddagger}$ & 
      $\tcc^\ast$ & ${\rm E_{w, \, 46}}^{\dagger}$ & ${\rm M_{ej}}^{\ddagger}$ & $v_{\rm esc}^{\ddagger}$ & ${\rm R_{ej}}^{\ddagger}$ & 
      $\tcc^\ast$ & ${\rm E_{w, \, 46}}^{\dagger}$ & ${\rm M_{ej}}^{\ddagger}$ & $v_{\rm esc}^{\ddagger}$ & ${\rm R_{ej}}^{\ddagger}$\\
      \\[-1.8ex]
      $\left[{\rm M}_{\odot}\right]$ & $\left[{\rm Z}_\odot\right]$ & & & & $\left[{\rm M}_{\odot}\right]$ & 
      $\left[{\rm yr}\right]$ & $\left[{\rm erg}\right]$ & $\left[{\rm M}_{\odot}\right]$ &  $\left[{\rm km/s}\right]$ & $\left[{\rm AU}\right]$ & 
      $\left[{\rm yr}\right]$ & $\left[{\rm erg}\right]$ & $\left[{\rm M}_{\odot}\right]$ &  $\left[{\rm km/s}\right]$ & $\left[{\rm AU}\right]$ & 
      $\left[{\rm day}\right]$ & $\left[{\rm erg}\right]$ & $\left[{\rm M}_{\odot}\right]$ & $\left[{\rm km/s}\right]$ & $\left[{\rm R}_\odot\right]$\\[0.5ex]
   \hline \\[-1.8ex]
   \endfirsthead
    
   \multicolumn{21}{c}{{\tablename} \thetable{} --- Continued} \\
   \hline \hline \\[-2ex]
      & & & & & & \multicolumn{5}{c}{\underline{core Ne burning}} & 
      \multicolumn{5}{c}{\underline{core O burning}} & \multicolumn{5}{c}{\underline{core Si burning}}\\
      \\[-1.8ex]
      ${\rm M_{ZAMS}}$ & ${\rm Z_{init}}$ & $\Omega/\Omega_{\rm crit}$ & Mass Loss & Class & ${\rm M_{He}}$ &
      $\tcc^\ast$ & ${\rm E_{w, \, 46}}^{\dagger}$ & ${\rm M_{ej}}^{\ddagger}$ & $v_{\rm esc}^{\ddagger}$ & ${\rm R_{ej}}^{\ddagger}$ & 
      $\tcc^\ast$ & ${\rm E_{w, \, 46}}^{\dagger}$ & ${\rm M_{ej}}^{\ddagger}$ & $v_{\rm esc}^{\ddagger}$ & ${\rm R_{ej}}^{\ddagger}$ & 
      $\tcc^\ast$ & ${\rm E_{w, \, 46}}^{\dagger}$ & ${\rm M_{ej}}^{\ddagger}$ & $v_{\rm esc}^{\ddagger}$ & ${\rm R_{ej}}^{\ddagger}$\\
      \\[-1.8ex]
      $\left[{\rm M}_{\odot}\right]$ & $\left[{\rm Z}_\odot\right]$ & & & & $\left[{\rm M}_{\odot}\right]$ & 
      $\left[{\rm yr}\right]$ & $\left[{\rm erg}\right]$ & $\left[{\rm M}_{\odot}\right]$ &  $\left[{\rm km/s}\right]$ & $\left[{\rm AU}\right]$ & 
      $\left[{\rm yr}\right]$ & $\left[{\rm erg}\right]$ & $\left[{\rm M}_{\odot}\right]$ &  $\left[{\rm km/s}\right]$ & $\left[{\rm AU}\right]$ & 
      $\left[{\rm day}\right]$ & $\left[{\rm erg}\right]$ & $\left[{\rm M}_{\odot}\right]$ & $\left[{\rm km/s}\right]$ & $\left[{\rm R}_\odot\right]$\\[0.5ex]
   \hline \\[-1.8ex]
   \endhead
    
       \\[-1.8ex]
        \hline \hline\\[-2ex]
        \multicolumn{21}{l}{Continued on Next Page\ldots}
   \endfoot
    
       \\[-1.8ex] 
       \hline \hline
	   \\
       \multicolumn{21}{l}{$^\ast$A value is given whenever burning occurs convectively, otherwise $-$ is given.}\\
       \multicolumn{21}{l}{$^{\dagger}$Wave energy excited in units of $10^{46} \erg$; a value is given whenever convectively excited waves meet the \ref{eqn:Lcond} and \ref{eqn:tcond} conditions, otherwise $-$ is given.}\\
       \multicolumn{21}{l}{$^{\ddagger}$A value is given whenever the \ref{eqn:Lcond}, \ref{eqn:tcond} and \ref{eqn:outflowcond} conditions are met, otherwise $-$ is given.}
   \endlastfoot
      $12$ & $0$ & $0.00$ & 0.8 (v+dj) & RSG & $3.17$ & $11.00$ & $3.3$ & $0.3$ & $100$ & $240$ & $6.300$ & $3.9$ & $-$ & $-$ & $-$ & $5.10$ & $12$ & $0.7$ & $130$ & $81$\\
      $12$ & $10^{-2}$ & $0.00$ & 0.8 (v+dj) & RSG & $4.39$ & $6.90$ & $2.8$ & $0.3$ & $92$ & $130$ & $4.400$ & $4.8$ & $-$ & $-$ & $-$ & $3.90$ & $12$ & $-$ & $-$ & $-$\\
      $12$ & $10^{-1}$ & $0.00$ & 0.8 (v+dj) & RSG & $4.35$ & $7.60$ & $3.0$ & $0.5$ & $79$ & $130$ & $4.400$ & $4.3$ & $-$ & $-$ & $-$ & $4.70$ & $11$ & $-$ & $-$ & $-$\\
      $12$ & $1.0$ & $0.00$ & 0.8 (v+dj) & RSG & $4.21$ & $9.20$ & $3.0$ & $0.7$ & $66$ & $130$ & $5.200$ & $4.0$ & $-$ & $-$ & $-$ & $5.90$ & $19$ & $-$ & $-$ & $-$\\
      $12$ & $1.0$ & $0.20$ & 0.6 (v+dj) & RSG & $4.94$ & $6.10$ & $1.9$ & $0.4$ & $68$ & $88$ & $2.900$ & $5.3$ & $-$ & $-$ & $-$ & $4.00$ & $10$ & $-$ & $-$ & $-$\\
      $12$ & $1.0$ & $0.50$ & 0.6 (v+dj) & RSG & $2.79$ & $9.20$ & $1.4$ & $0.3$ & $65$ & $130$ & $3.700$ & $5.5$ & $-$ & $-$ & $-$ & $3.90$ & $11$ & $5.6 \ 10^{-2}$ & $440$ & $210$\\
      $12$ & $1.0$ & $0.80$ & 0.6 (v+dj) & RSG & $6.57$ & $3.90$ & $2.5$ & $0.6$ & $67$ & $56$ & $2.700$ & $7.1$ & $-$ & $-$ & $-$ & $3.00$ & $13$ & $-$ & $-$ & $-$\\[0.5ex]
      \hline\\[-2ex]
      $15$ & $0$ & $0.00$ & 0.8 (v+dj) & BSG & $3.62$ & $5.10$ & $3.2$ & $3.5 \ 10^{-2}$ & $300$ & $330$ & $2.900$ & $6.4$ & $-$ & $-$ & $-$ & $3.00$ & $12$ & $-$ & $-$ & $-$\\
      $15$ & $10^{-2}$ & $0.00$ & 0.8 (v+dj) & RSG & $5.84$ & $3.80$ & $2.0$ & $0.3$ & $89$ & $71$ & $2.000$ & $7.4$ & $-$ & $-$ & $-$ & $2.80$ & $11$ & $-$ & $-$ & $-$\\
      $15$ & $10^{-1}$ & $0.00$ & 0.8 (v+dj) & RSG & $5.82$ & $5.30$ & $1.6$ & $0.3$ & $74$ & $82$ & $2.300$ & $8.8$ & $-$ & $-$ & $-$ & $2.50$ & $14$ & $-$ & $-$ & $-$\\
      $15$ & $1.0$ & $0.00$ & 0.8 (v+dj) & RSG & $5.72$ & $4.60$ & $1.9$ & $0.5$ & $62$ & $60$ & $2.200$ & $7.4$ & $1.9$ & $63$ & $29$ & $3.30$ & $15$ & $-$ & $-$ & $-$\\
      $15$ & $1.0$ & $0.50$ & 0.6 (v+dj) & RSG & $6.69$ & $4.90$ & $1.7$ & $0.4$ & $66$ & $68$ & $2.500$ & $5.8$ & $-$ & $-$ & $-$ & $2.30$ & $14$ & $-$ & $-$ & $-$\\
      $15$ & $1.0$ & $0.80$ & 0.6 (v+dj) & RSG & $9.48$ & $1.10$ & $1.0$ & $0.1$ & $92$ & $22$ & $0.560$ & $17.0$ & $0.5$ & $180$ & $22$ & $1.00$ & $31$ & $-$ & $-$ & $-$\\[0.5ex]
      \hline\\[-2ex]
      $20$ & $0$ & $0.00$ & 0.8 (v+dj) & BSG & $5.24$ & $2.10$ & $2.2$ & $9.7 \ 10^{-3}$ & $480$ & $210$ & $1.200$ & $8.9$ & $-$ & $-$ & $-$ & $3.70$ & $13$ & $-$ & $-$ & $-$\\
      $20$ & $10^{-2}$ & $0.00$ & 0.8 (v+dj) & RSG & $7.78$ & $0.65$ & $3.2$ & $-$ & $-$ & $-$ & $0.440$ & $16.0$ & $-$ & $-$ & $-$ & $1.30$ & $20$ & $-$ & $-$ & $-$\\
      $20$ & $10^{-1}$ & $0.00$ & 0.8 (v+dj) & RSG & $8.49$ & $2.20$ & $1.5$ & $-$ & $-$ & $-$ & $1.300$ & $8.1$ & $-$ & $-$ & $-$ & $2.50$ & $12$ & $-$ & $-$ & $-$\\
      $20$ & $1.0$ & $0.00$ & 0.8 (v+dj) & RSG & $8.45$ & $3.20$ & $1.9$ & $0.4$ & $71$ & $48$ & $1.600$ & $10.0$ & $1.3$ & $90$ & $31$ & $1.90$ & $21$ & $-$ & $-$ & $-$\\
      $20$ & $1.0$ & $0.20$ & 0.6 (v+dj) & RSG & $7.70$ & $1.80$ & $1.8$ & $-$ & $-$ & $-$ & $0.980$ & $12.0$ & $-$ & $-$ & $-$ & $1.80$ & $16$ & $-$ & $-$ & $-$\\
      $20$ & $1.0$ & $0.50$ & 0.6 (v+dj) & RSG & $9.70$ & $0.96$ & $2.4$ & $0.4$ & $78$ & $16$ & $0.600$ & $-$ & $-$ & $-$ & $-$ & $2.40$ & $16$ & $-$ & $-$ & $-$\\
      $20$ & $1.0$ & $0.80$ & 0.6 (v+dj) & WR & $10.96$ & $0.88$ & $-$ & $-$ & $-$ & $-$ & $0.350$ & $23.0$ & $4.8 \ 10^{-3}$ & $2200$ & $160$ & $0.76$ & $29$ & $5.9 \ 10^{-3}$ & $2200$ & $210$\\[0.5ex]
      \hline\\[-2ex]
      $25$ & $0$ & $0.00$ & 0.8 (v+dj) & BSG & $9.41$ & $0.51$ & $3.4$ & $-$ & $-$ & $-$ & $0.310$ & $-$ & $-$ & $-$ & $-$ & $1.50$ & $18$ & $-$ & $-$ & $-$\\
      $25$ & $10^{-2}$ & $0.00$ & 0.8 (v+dj) & RSG & $11.05$ & $0.60$ & $2.0$ & $-$ & $-$ & $-$ & $0.310$ & $25.0$ & $-$ & $-$ & $-$ & $0.66$ & $29$ & $-$ & $-$ & $-$\\
      $25$ & $10^{-1}$ & $0.00$ & 0.8 (v+dj) & RSG & $11.34$ & $0.34$ & $3.8$ & $-$ & $-$ & $-$ & $0.260$ & $24.0$ & $-$ & $-$ & $-$ & $1.10$ & $24$ & $-$ & $-$ & $-$\\
      $25$ & $1.0$ & $0.00$ & 0.8 (v+dj) & RSG & $11.16$ & $0.27$ & $3.9$ & $0.3$ & $120$ & $6$ & $0.170$ & $36.0$ & $-$ & $-$ & $-$ & $0.74$ & $25$ & $-$ & $-$ & $-$\\
      $25$ & $1.0$ & $0.20$ & 0.6 (v+dj) & RSG & $12.39$ & $1.10$ & $1.4$ & $0.1$ & $110$ & $24$ & $0.480$ & $17.0$ & $-$ & $-$ & $-$ & $1.00$ & $30$ & $-$ & $-$ & $-$\\
      $25$ & $1.0$ & $0.50$ & 0.6 (v+dj) & RSG & $13.07$ & $0.90$ & $1.4$ & $-$ & $-$ & $-$ & $0.460$ & $14.0$ & $-$ & $-$ & $-$ & $1.00$ & $29$ & $-$ & $-$ & $-$\\
      $25$ & $1.0$ & $0.80$ & 0.6 (v+dj) & WR & $11.36$ & $0.99$ & $-$ & $-$ & $-$ & $-$ & $0.330$ & $-$ & $-$ & $-$ & $-$ & $0.87$ & $26$ & $2.9 \ 10^{-3}$ & $3000$ & $320$\\[0.5ex]
      \hline\\[-2ex]
      $30$ & $0$ & $0.00$ & 0.8 (v+dj) & BSG & $11.90$ & $0.77$ & $0.6$ & $-$ & $-$ & $-$ & $0.410$ & $20.0$ & $-$ & $-$ & $-$ & $0.99$ & $27$ & $-$ & $-$ & $-$\\
      $30$ & $10^{-2}$ & $0.00$ & 0.8 (v+dj) & RSG & $13.98$ & $0.96$ & $1.4$ & $-$ & $-$ & $-$ & $0.520$ & $-$ & $-$ & $-$ & $-$ & $2.10$ & $17$ & $-$ & $-$ & $-$\\
      $30$ & $10^{-1}$ & $0.00$ & 0.8 (v+dj) & RSG & $14.28$ & $0.71$ & $-$ & $-$ & $-$ & $-$ & $0.390$ & $25.0$ & $-$ & $-$ & $-$ & $1.20$ & $25$ & $-$ & $-$ & $-$\\
      $30$ & $1.0$ & $0.00$ & 0.8 (v+dj) & WR & $13.86$ & $0.73$ & $-$ & $-$ & $-$ & $-$ & $0.390$ & $-$ & $-$ & $-$ & $-$ & $1.40$ & $20$ & $5.1 \ 10^{-3}$ & $2000$ & $330$\\
      $30$ & $1.0$ & $0.20$ & 0.6 (v+dj) & BSG & $15.46$ & $0.33$ & $-$ & $-$ & $-$ & $-$ & $0.190$ & $-$ & $-$ & $-$ & $-$ & $1.00$ & $30$ & $-$ & $-$ & $-$\\
      $30$ & $1.0$ & $0.50$ & 0.6 (v+dj) & BSG & $16.33$ & $-$ & $-$ & $-$ & $-$ & $-$ & $0.160$ & $-$ & $-$ & $-$ & $-$ & $0.86$ & $28$ & $-$ & $-$ & $-$\\
      $30$ & $1.0$ & $0.80$ & 0.6 (v+dj) & WR & $12.65$ & $1.30$ & $-$ & $-$ & $-$ & $-$ & $0.670$ & $-$ & $-$ & $-$ & $-$ & $2.70$ & $15$ & $2.4 \ 10^{-3}$ & $2500$ & $830$\\[0.5ex]
      \hline\\[-2ex]
      $40$ & $0$ & $0.00$ & 0.8 (v+dj) & RSG & $17.48$ & $-$ & $-$ & $-$ & $-$ & $-$ & $0.095$ & $46.0$ & $-$ & $-$ & $-$ & $0.78$ & $26$ & $-$ & $-$ & $-$\\
      $40$ & $10^{-2}$ & $0.00$ & 0.8 (v+dj) & RSG & $20.02$ & $-$ & $-$ & $-$ & $-$ & $-$ & $0.092$ & $44.0$ & $0.5$ & $300$ & $5$ & $0.68$ & $28$ & $-$ & $-$ & $-$\\
      $40$ & $10^{-1}$ & $0.00$ & 0.8 (v+dj) & BSG & $20.37$ & $-$ & $-$ & $-$ & $-$ & $-$ & $0.092$ & $46.0$ & $0.4$ & $350$ & $6$ & $0.71$ & $29$ & $-$ & $-$ & $-$\\
      $40$ & $1.0$ & $0.00$ & v+n & WR & $10.04$ & $0.34$ & $-$ & $-$ & $-$ & $-$ & $0.180$ & $-$ & $-$ & $-$ & $-$ & $0.87$ & $25$ & $2.5 \ 10^{-3}$ & $3200$ & $340$\\
      $40$ & $1.0$ & $0.00$ & 0.8 (v+dj) & WR & $13.10$ & $0.73$ & $-$ & $-$ & $-$ & $-$ & $0.370$ & $-$ & $-$ & $-$ & $-$ & $1.70$ & $16$ & $1.6 \ 10^{-3}$ & $3100$ & $680$\\
      $40$ & $1.0$ & $0.20$ & 0.6 (v+dj) & WR & $20.41$ & $-$ & $-$ & $-$ & $-$ & $-$ & $0.063$ & $65.0$ & $7.9 \ 10^{-3}$ & $2900$ & $38$ & $1.00$ & $30$ & $3.6 \ 10^{-3}$ & $2900$ & $370$\\
      $40$ & $1.0$ & $0.50$ & 0.6 (v+dj) & WR & $20.01$ & $-$ & $-$ & $-$ & $-$ & $-$ & $0.065$ & $-$ & $-$ & $-$ & $-$ & $0.98$ & $27$ & $2.7 \ 10^{-3}$ & $3200$ & $380$\\
      $40$ & $1.0$ & $0.80$ & 0.6 (v+dj) & WR & $13.77$ & $0.59$ & $-$ & $-$ & $-$ & $-$ & $0.300$ & $-$ & $-$ & $-$ & $-$ & $1.60$ & $22$ & $1.7 \ 10^{-3}$ & $3600$ & $740$\\[0.5ex]
      \hline\\[-2ex]
      $50$ & $0$ & $0.00$ & 0.8 (v+dj) & RSG & $23.13$ & $-$ & $-$ & $-$ & $-$ & $-$ & $0.040$ & $83.0$ & $-$ & $-$ & $-$ & $0.86$ & $33$ & $-$ & $-$ & $-$\\
      $50$ & $10^{-2}$ & $0.00$ & 0.8 (v+dj) & BSG & $26.28$ & $-$ & $-$ & $-$ & $-$ & $-$ & $0.042$ & $82.0$ & $-$ & $-$ & $-$ & $0.90$ & $31$ & $-$ & $-$ & $-$\\
      $50$ & $10^{-1}$ & $0.00$ & 0.8 (v+dj) & BSG & $26.61$ & $-$ & $-$ & $-$ & $-$ & $-$ & $0.043$ & $84.0$ & $0.5$ & $430$ & $3$ & $0.91$ & $31$ & $-$ & $-$ & $-$\\
      $50$ & $1.0$ & $0.00$ & 0.8 (v+dj) & WR & $12.35$ & $0.96$ & $-$ & $-$ & $-$ & $-$ & $0.500$ & $-$ & $-$ & $-$ & $-$ & $2.50$ & $16$ & $1.7 \ 10^{-3}$ & $3100$ & $930$\\
      $50$ & $1.0$ & $0.00$ & v+n & WR & $10.42$ & $0.78$ & $-$ & $-$ & $-$ & $-$ & $0.280$ & $-$ & $-$ & $-$ & $-$ & $0.73$ & $25$ & $2.2 \ 10^{-3}$ & $3400$ & $310$\\
      $50$ & $1.0$ & $0.20$ & 0.6 (v+dj) & WR & $24.19$ & $-$ & $-$ & $-$ & $-$ & $-$ & $0.034$ & $-$ & $-$ & $-$ & $-$ & $0.87$ & $-$ & $-$ & $-$ & $-$\\
      $50$ & $1.0$ & $0.50$ & 0.6 (v+dj) & WR & $22.33$ & $-$ & $-$ & $-$ & $-$ & $-$ & $0.041$ & $-$ & $-$ & $-$ & $-$ & $0.83$ & $-$ & $-$ & $-$ & $-$\\
      $50$ & $1.0$ & $0.80$ & 0.6 (v+dj) & WR & $14.36$ & $-$ & $-$ & $-$ & $-$ & $-$ & $0.230$ & $-$ & $-$ & $-$ & $-$ & $1.40$ & $25$ & $1.9 \ 10^{-3}$ & $3600$ & $630$\\[0.5ex]
      \hline\\[-2ex]
      $60$ & $0$ & $0.00$ & 0.8 (v+dj) & BSG & $29.15$ & $-$ & $-$ & $-$ & $-$ & $-$ & $0.024$ & $150.0$ & $-$ & $-$ & $-$ & $0.35$ & $59$ & $-$ & $-$ & $-$\\
      $60$ & $10^{-2}$ & $0.00$ & 0.8 (v+dj) & BSG & $32.60$ & $-$ & $-$ & $-$ & $-$ & $-$ & $0.023$ & $140.0$ & $-$ & $-$ & $-$ & $0.59$ & $56$ & $-$ & $-$ & $-$\\
      $60$ & $10^{-1}$ & $0.00$ & 0.8 (v+dj) & BSG & $32.75$ & $-$ & $-$ & $-$ & $-$ & $-$ & $0.021$ & $150.0$ & $9.2 \ 10^{-2}$ & $1300$ & $5$ & $0.41$ & $46$ & $-$ & $-$ & $-$\\
      $60$ & $1.0$ & $0.00$ & v+n & WR & $11.70$ & $1.20$ & $-$ & $-$ & $-$ & $-$ & $0.640$ & $-$ & $-$ & $-$ & $-$ & $2.60$ & $14$ & $1.6 \ 10^{-3}$ & $3000$ & $950$\\
      $60$ & $1.0$ & $0.00$ & 0.8 (v+dj) & WR & $15.14$ & $-$ & $-$ & $-$ & $-$ & $-$ & $0.160$ & $-$ & $-$ & $-$ & $-$ & $0.98$ & $24$ & $1.9 \ 10^{-3}$ & $3600$ & $440$\\
      $60$ & $1.0$ & $0.20$ & 0.6 (v+dj) & WR & $26.93$ & $-$ & $-$ & $-$ & $-$ & $-$ & $0.024$ & $-$ & $-$ & $-$ & $-$ & $0.45$ & $-$ & $-$ & $-$ & $-$\\
      $60$ & $1.0$ & $0.50$ & 0.6 (v+dj) & WR & $24.56$ & $-$ & $-$ & $-$ & $-$ & $-$ & $0.035$ & $-$ & $-$ & $-$ & $-$ & $0.68$ & $-$ & $-$ & $-$ & $-$\\
      $60$ & $1.0$ & $0.80$ & 0.6 (v+dj) & WR & $15.67$ & $-$ & $-$ & $-$ & $-$ & $-$ & $0.160$ & $-$ & $-$ & $-$ & $-$ & $0.89$ & $33$ & $2.1 \ 10^{-3}$ & $3900$ & $430$\\[0.5ex]
      \hline\\[-2ex]
      $70$ & $0$ & $0.00$ & 0.8 (v+dj) & BSG & $34.99$ & $-$ & $-$ & $-$ & $-$ & $-$ & $0.015$ & $200.0$ & $-$ & $-$ & $-$ & $0.29$ & $82$ & $-$ & $-$ & $-$\\
      $70$ & $10^{-2}$ & $0.00$ & 0.8 (v+dj) & BSG & $38.95$ & $-$ & $-$ & $-$ & $-$ & $-$ & $0.015$ & $190.0$ & $-$ & $-$ & $-$ & $0.29$ & $71$ & $-$ & $-$ & $-$\\
      $70$ & $10^{-1}$ & $0.00$ & 0.8 (v+dj) & BSG & $38.97$ & $-$ & $-$ & $-$ & $-$ & $-$ & $0.016$ & $200.0$ & $2.6 \ 10^{-2}$ & $2800$ & $9$ & $0.30$ & $66$ & $7.3 \ 10^{-3}$ & $3000$ & $110$\\
      $70$ & $1.0$ & $0.00$ & v+n & WR & $13.54$ & $0.50$ & $-$ & $-$ & $-$ & $-$ & $0.250$ & $-$ & $-$ & $-$ & $-$ & $1.40$ & $22$ & $1.9 \ 10^{-3}$ & $3400$ & $590$\\
      $70$ & $1.0$ & $0.00$ & 0.8 (v+dj) & WR & $17.45$ & $-$ & $-$ & $-$ & $-$ & $-$ & $0.096$ & $-$ & $-$ & $-$ & $-$ & $0.72$ & $25$ & $1.5 \ 10^{-3}$ & $4000$ & $360$\\
      $70$ & $1.0$ & $0.20$ & 0.6 (v+dj) & WR & $24.21$ & $-$ & $-$ & $-$ & $-$ & $-$ & $0.039$ & $-$ & $-$ & $-$ & $-$ & $1.00$ & $-$ & $-$ & $-$ & $-$\\
      $70$ & $1.0$ & $0.50$ & 0.6 (v+dj) & WR & $25.68$ & $-$ & $-$ & $-$ & $-$ & $-$ & $0.033$ & $-$ & $-$ & $-$ & $-$ & $0.43$ & $-$ & $-$ & $-$ & $-$\\
      $70$ & $1.0$ & $0.80$ & 0.6 (v+dj) & WR & $16.43$ & $-$ & $-$ & $-$ & $-$ & $-$ & $0.140$ & $-$ & $-$ & $-$ & $-$ & $0.95$ & $32$ & $2.2 \ 10^{-3}$ & $3900$ & $460$\\[0.5ex]
      \hline\\[-2ex]
      $80$ & $1.0$ & $0.00$ & 0.8 (v+dj) & WR & $18.89$ & $-$ & $-$ & $-$ & $-$ & $-$ & $0.072$ & $-$ & $-$ & $-$ & $-$ & $0.92$ & $22$ & $1.4 \ 10^{-3}$ & $4000$ & $450$\\
      $80$ & $1.0$ & $0.00$ & v+n & WR & $8.45$ & $3.00$ & $-$ & $-$ & $-$ & $-$ & $1.600$ & $-$ & $-$ & $-$ & $-$ & $2.30$ & $14$ & $2.0 \ 10^{-3}$ & $2600$ & $770$\\
      $80$ & $1.0$ & $0.20$ & 0.6 (v+dj) & WR & $19.83$ & $-$ & $-$ & $-$ & $-$ & $-$ & $0.073$ & $-$ & $-$ & $-$ & $-$ & $0.80$ & $26$ & $1.5 \ 10^{-3}$ & $4200$ & $410$\\
      $80$ & $1.0$ & $0.50$ & 0.6 (v+dj) & WR & $18.61$ & $-$ & $-$ & $-$ & $-$ & $-$ & $0.085$ & $-$ & $-$ & $-$ & $-$ & $0.71$ & $32$ & $2.0 \ 10^{-3}$ & $4000$ & $360$\\
      $80$ & $1.0$ & $0.80$ & 0.6 (v+dj) & WR & $17.54$ & $-$ & $-$ & $-$ & $-$ & $-$ & $0.110$ & $-$ & $-$ & $-$ & $-$ & $0.82$ & $28$ & $1.8 \ 10^{-3}$ & $3900$ & $390$\\[0.5ex]
      \hline\\[-2ex]
      $100$ & $1.0$ & $0.00$ & 0.8 (v+dj) & WR & $9.81$ & $0.33$ & $-$ & $-$ & $-$ & $-$ & $0.220$ & $-$ & $-$ & $-$ & $-$ & $1.10$ & $24$ & $2.7 \ 10^{-3}$ & $3000$ & $410$\\
      $100$ & $1.0$ & $0.00$ & v+n & WR & $6.34$ & $1.40$ & $-$ & $-$ & $-$ & $-$ & $0.600$ & $15.0$ & $3.6 \ 10^{-3}$ & $2000$ & $260$ & $1.70$ & $17$ & $3.7 \ 10^{-3}$ & $2200$ & $450$\\
      $100$ & $1.0$ & $0.20$ & 0.6 (v+dj) & WR & $19.01$ & $-$ & $-$ & $-$ & $-$ & $-$ & $0.087$ & $-$ & $-$ & $-$ & $-$ & $0.73$ & $29$ & $1.7 \ 10^{-3}$ & $4100$ & $370$\\
      $100$ & $1.0$ & $0.50$ & 0.6 (v+dj) & WR & $18.06$ & $-$ & $-$ & $-$ & $-$ & $-$ & $0.100$ & $-$ & $-$ & $-$ & $-$ & $0.70$ & $31$ & $1.9 \ 10^{-3}$ & $4100$ & $350$\\
      $100$ & $1.0$ & $0.80$ & 0.6 (v+dj) & WR & $17.47$ & $-$ & $-$ & $-$ & $-$ & $-$ & $0.110$ & $-$ & $-$ & $-$ & $-$ & $0.87$ & $29$ & $1.9 \ 10^{-3}$ & $3900$ & $430$\\
\end{longtable}

\end{center}



\setlength{\tabcolsep}{6pt}

	\clearpage
\end{landscape}

\footnotesize{
	\bibliographystyle{apj}
	\bibliography{jhsrefs}
}


\end{document}